\documentclass[twocolumn,superscriptaddress,preprintnumbers,prb,amsmath,amssymb]{revtex4}
\usepackage{graphicx}
\usepackage{bm}
\usepackage{epsfig}

\newcommand{\Sr}{SrFe$_2$As$_2$}

\newcommand{\Ca}{CaFe$_2$As$_2$}
\newcommand{\CaR}{Ca$_{1-x}$R$_x$Fe$_2$As$_2$}
\newcommand{\LaCa}{Ca$_{1-x}$La$_x$Fe$_2$As$_2$}

\newcommand{\PrCa}{Ca$_{1-x}$Pr$_x$Fe$_2$As$_2$}
\newcommand{\NdCa}{Ca$_{1-x}$Nd$_x$Fe$_2$As$_2$}

\newcommand{\tc}{$T_c$}

\newcommand{\ie}{{\it i.e.}}
\newcommand{\eg}{{\it e.g.}}
\newcommand{\etal}{{\it et al.}}

\begin{document}

%%%%%%%%%%%%%%%%%%%%% TITLE %%%%%%%%%%%%%%%%%%%%

\title{Structural collapse and superconductivity in rare earth-doped CaFe$_2$As$_2$}

%%%%%%%%%%%%%%%%%%%% AUTHORS %%%%%%%%%%%%%%%%%%

\author{S.~R.~Saha}
\author{N.~P.~Butch}
\author{T.~Drye}
\author{J.~Magill}
\author{S.~Ziemak}
\author{K.~Kirshenbaum}
\affiliation{Center for Nanophysics and Advanced Materials, Department of Physics, University of Maryland, College Park, MD 20742}

\author{P.~Y.~Zavalij}
\affiliation{Department of Chemistry and Biochemistry, University of Maryland, College Park, MD 20742}

\author{J.~W.~Lynn}
\affiliation{NIST Center for Neutron Research, National Institute of Standards and Technology, Gaithersburg, MD 20899}

\author{J.~Paglione}
\email{paglione@umd.edu}
\affiliation{Center for Nanophysics and Advanced Materials, Department of Physics, University of Maryland, College Park, MD 20742}

%%%%%%%%%%%%%%%%%%% ABSTRACT %%%%%%%%%%%%%%%%%%%%

\begin{abstract}

Aliovalent rare earth substitution into the alkaline earth site of CaFe$_2$As$_2$ single-crystals is used to fine-tune structural, magnetic and electronic properties of this iron-based superconducting system. Neutron and single crystal x-ray scattering experiments indicate that an isostructural collapse of the tetragonal unit cell can be controllably induced at ambient pressures by choice of substituent ion size. This instability is driven by the interlayer As-As anion separation, resulting in an unprecedented thermal expansion coefficient of $180\times 10^{-6}$~K$^{-1}$. Electrical transport and magnetic susceptibility measurements reveal abrupt changes in the physical properties through the collapse as a function of temperature, including a reconstruction of the electronic structure. Superconductivity with onset transition temperatures as high as 47~K is stabilized by the suppression of antiferromagnetic order via chemical pressure, electron doping or a combination of both. Extensive investigations are performed to understand the observations of partial volume-fraction diamagnetic screening, ruling out extrinsic sources such as strain mechanisms, surface states or foreign phases as the cause of this superconducting phase that appears to be stable in both collapsed and uncollapsed structures.

\end{abstract}

%\pacs{}

\date{\today}

\maketitle

%%%%%%%%%%%%%%%%%%%%%%%%%%%%%%%%%%%%%%%%%%%%%%%%%%%%%%%%%%%%%%%%%%%%%%%%%%%%%%%%%%
\section{Introduction}

The interplay between structural, magnetic and superconducting properties in the iron-based superconducting compounds has been a central theme in attempts to elucidate the nature of Cooper pairing in this new family of high-temperature superconductors.\cite{Kamihara,Paglione} Much focus has been paid to the intermetallic series of iron-based compounds with the ThCr$_2$Si$_2$-type (122) crystal structure. 
With well over 700 compounds known to take on this configuration,\cite{Just} the 122 structure forms the basis for a rich variety of physical phenomena that stem from the fact that this lattice structure not only supports a wide assortment of elemental combinations, but also harbors different mixtures of ionic, covalent and metallic bonding. 

The interesting chemistry of the AB$_2$X$_2$ configuration was highlighted in 1985 by R.~Hoffman,\cite{Hoffman} who pointed out that a segregation of this large family of materials occurs due to the presence or absence of interlayer X-X bonding, which results in, respectively, either a ``collapsed'' or ``uncollapsed'' tetragonal structure. 
Despite a $\sim 20\%$ change in unit cell volume in traversing through Ba, Sr, and Ca-based series of AFe$_2$As$_2$ structures,\cite{SahaSCES} this family of 122 materials remains in the uncollapsed structure geometry at ambient pressures through the entire range. However, it can be driven to collapse by a modest external pressure applied to the smallest-volume member of the series, \Ca, resulting in a dramatically abrupt $\sim 10\%$ $c$-axis reduction of its tetragonal unit cell upon cooling.\cite{Kreyssig,Goldman} 

Interestingly, a superconducting phase with maximum \tc\ of $\sim 10$~K was first reported to straddle the critical pressure where the collapse occurs in \Ca\ (Refs.~\onlinecite{Torikachvili,Park,Prokes}) but only under non-hydrostatic experimental conditions,\cite{Yu} suggesting that the change in electronic structure that occurs through the collapse is not supportive of pairing. This was further confirmed by the absence of superconductivity in the collapsed phase induced by isovalent phosphorus substitution,\cite{Kasahara} and its importance heightened by first principles calculations predicting a quenched Fe moment in the collapsed structure.\cite{Yildirim}

The nearness of the structural collapse instability to the ambient pressure phase of \Ca\ suggests that chemical pressure is a viable alternative to applied pressure, as indeed was shown by isovalent substitution directly into the FeAs sublattice.\cite{Kasahara}
We have stabilized the 122 collapsed phase at ambient pressures by employing rare earth substitution into \CaR\ to selectively induce a structural collapse via choice of rare earth.\cite{Paglione_APS} The close match between ionic radii of the lighter rare earths such as La, Ce, Pr and Nd (130, 128.3, 126.6 and 124.9 pm, respectively \cite{Shannon}) with that of Ca (126 pm) in the 8-coordinate geometry allows us to selectively tune the structural parameters with both larger and smaller relative radii, invoking a uniform chemical pressure on the unit cell at a tunable rate of chemical substitution.
In parallel, electron doping via aliovalent substitution of trivalent R$^{3+}$ ions for divalent Ca$^{2+}$ also tunes the electronic structure, acting to suppress antiferromagnetic (AFM) order and induce a superconducting phase with transition temperatures reaching as high as 47~K. 

In this article, we provide a comprehensive study of the structural, magnetic and electronic properties of the rare earth-doped \Ca\ system throughout the antiferromagnetic, superconducting and structural collapsed phases. We show that the substitution of light rare earths provides an ideal method of fine-tuning the structure through collapse, while simultaneously electron-doping the system.
In Sections II and III, we review the synthesis and characterization of our single-crystal samples, providing chemical analysis and structural refinements as a function of substitution and temperature. Neutron scattering and x-ray scattering experiments are used to outline the systematic chemical pressure effects of rare earth substitution and the evolution of all structural parameters through the collapse transition. 
Section IV presents the temperature dependence of electrical transport and magnetic susceptibility data, comparing the evolution of physical properties for different chemical pressures and dopings.
Section V reviews our systematic studies of the superconducting phase induced by the suppression of the antiferromagnetic phase, including annealing, etching, and oxidation effects. 
In Section VI we build a composite phase diagram that segregates the effects of electron doping and chemical pressure, and finally summarize our conclusions based on this work in Section VII.

%%%%%%%%%%%%%%%%%%%%%%%%%%%%%%%%%%%%%%%%%%%%%%%%%%%%%%%%%%%%%%%%%%%%%%%%%%%%%%%%%%
\section{Growth and Chemical Analysis}

%%%%%%%%%% FIGURE:  WDS
\begin{figure}[!t] \centering
  \resizebox{8cm}{!}{
  \includegraphics[width=8cm]{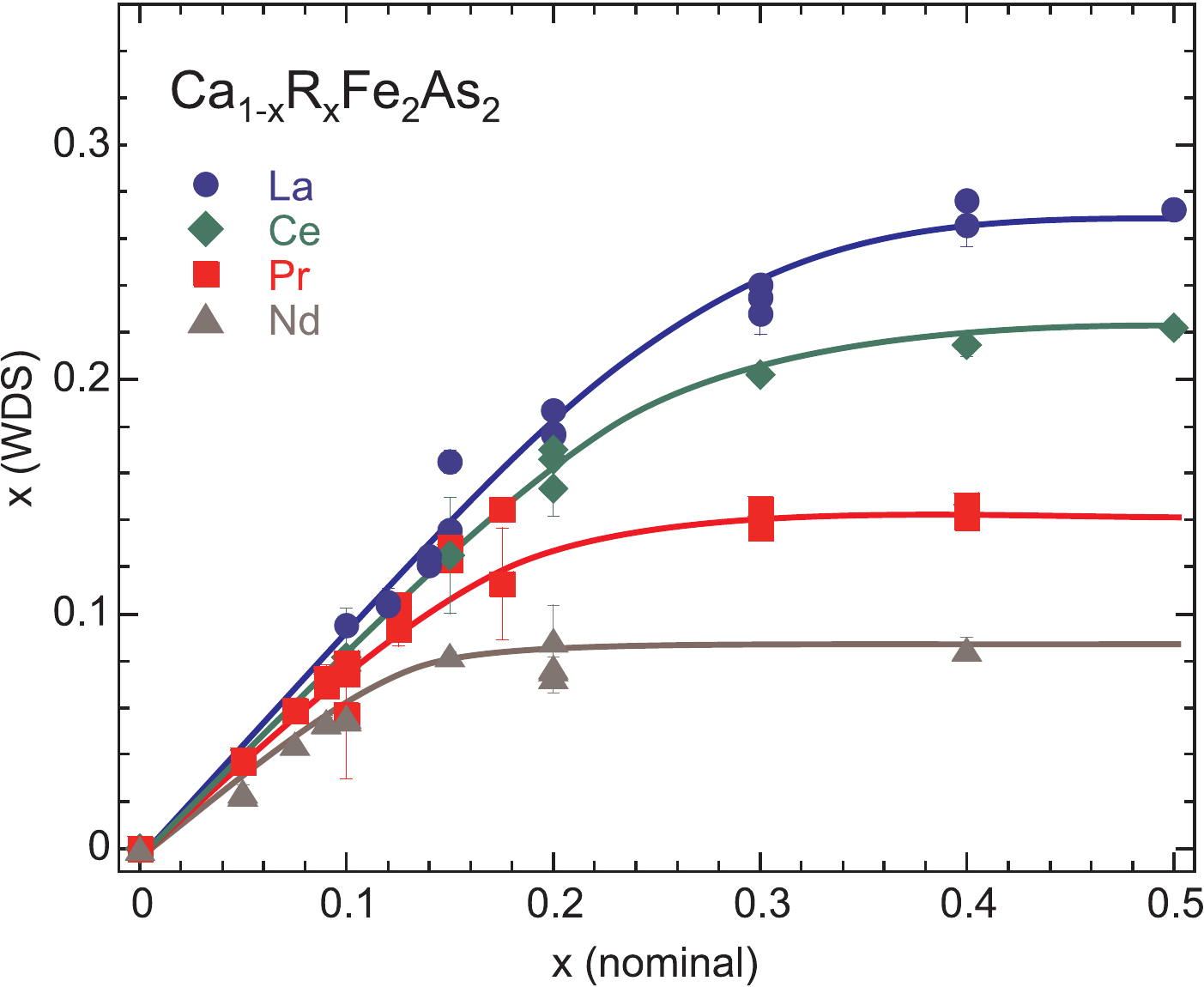}}
  \caption{\label{WDS} Chemical analysis of rare earth concentration $x$ in \CaR\ obtained from wavelength-dispersive (WDS) x-ray spectroscopy. Data points were determined by averaging 8 separate scanned WDS measurements across each sample, and error bars express the range of values determined in each measurement. Solid lines are guides to the eye.}
\end{figure}

Single-crystal samples of \CaR\ were grown using the FeAs self-flux method \cite{Saha_Sr},yielding crystals as large as $\sim 10 \times 10 \times 0.1$~mm$^3$.  Chemical analysis was obtained via both energy-dispersive (EDS) and wavelength-dispersive (WDS) x-ray spectroscopy, showing 1:2:2 stoichiometry between (Ca,R), Fe, and As concentrations. For WDS analysis, rare earth concentrations were determined by recording the concentration at 8 separate scanned points across each sample surface and averaging the result for each sample, with standard deviation never found to be more than a few percent (or within the accuracy of the technique). In the rest of the paper, actual $x$ concentrations are quoted based on WDS results.

%%%%%%%%%% FIGURE:  xray1 - lattice pars
\begin{figure} \centering
  \resizebox{8cm}{!}{
  \includegraphics[width=8cm]{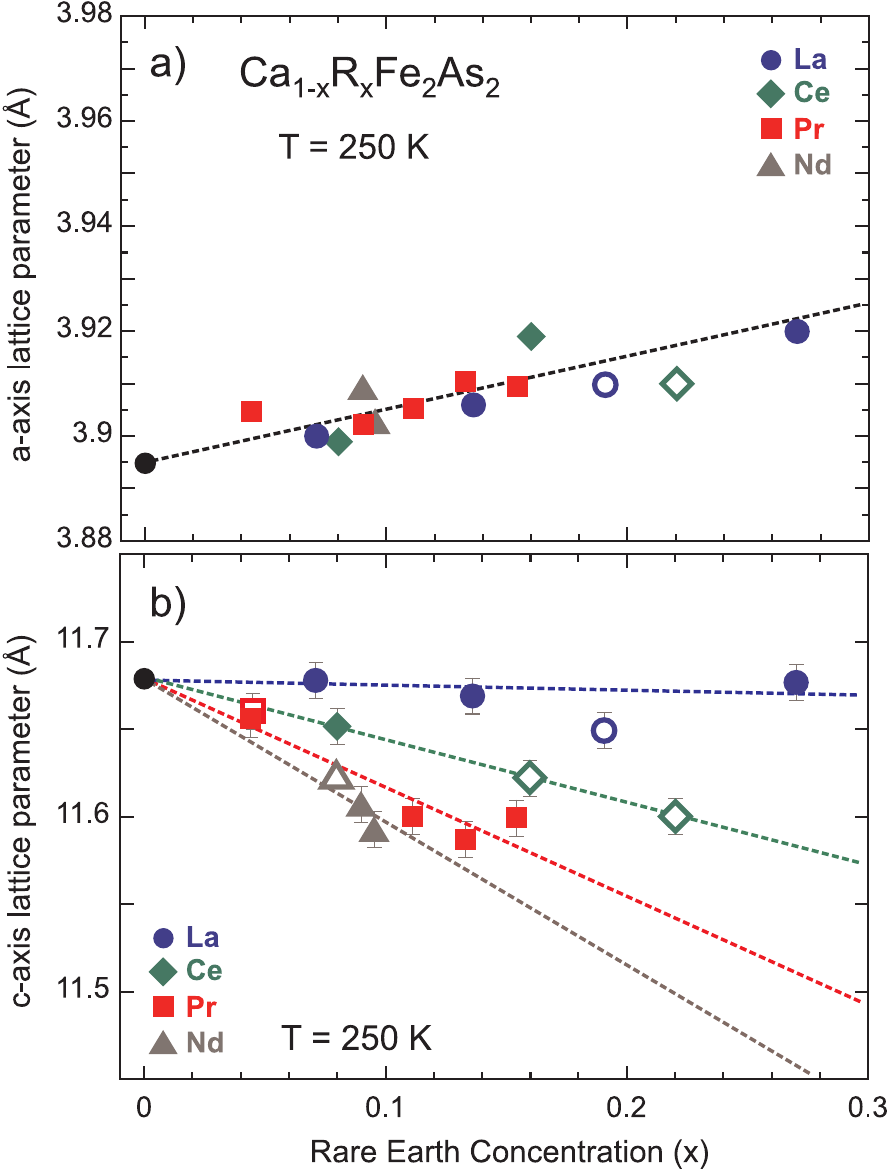}}
  \caption{\label{xray1} Characterization of \CaR\ unit cell lattice constants determined from single-crystal samples. Filled symbols correspond to x-ray diffraction data while open symbols are determined by neutron diffraction. Due to the large thermal expansion (see text), uncertainties are dominated by the temperature stability of the measurement apparatus estimated to be $250\pm 5$~K, resulting in $a$-axis error values in panel a) within the symbol sizes and $c$-axis error values as shown in panel b).
  The dashed lines are least-squares fits to data for a) all data, and b) each rare earth species.}
\end{figure}

Fig.~\ref{WDS} presents the measured concentration $x$ in \CaR\ single-crystal samples determined by WDS spectroscopy as a function of nominal starting concentration. Error bars are determined conservatively, taking the range in $x$ values determined in scanned WDS values across each sample. At low values of $x$ there is close to a one-to-one correspondence between actual and nominal concentrations, indicating good control of target and resultant substitution concentration. With increasing $x$, it is apparent that saturation occurs whereby actual substitutional concentrations do not exceed a solubility limit that depends on the rare earth species. This value can be readily inferred from the value at which $x$(WDS) saturates in Fig.~\ref{WDS}, and appears to follow the tendency of lower saturation point for decreasing ionic radii of the heavier rare earths. The limited solubility is likely due to ionic size mismatch between Ca$^{2+}$ and corresponding rare earth $R$, as also encountered in the synthesis of Sr$_{1-x}$La$_x$Fe$_2$As$_2$.\cite{Muraba}.

%%%%%%%%%%%%%%%%%%%%%%%%%%%%%%%%%%%%%%%%%%%%%%%%%%%%%%%%%%%%%%%%%%%%%%%%%%%%%%%%%%
\section{Structural Characterization}

%**x-ray
Crystal structure determination was performed using single-crystal x-ray diffraction data measured on a Bruker Smart Apex2 diffractometer equipped with CCD detector, graphite monochromator, and mono-cap collimator using MoK$\alpha$ radiation from fine focus sealed tube. Due to strong absorption and highly anisotropic crystal shape, the full sphere of reflections was collected with a redundancy of at least 8 and corrected for absorption effects using integration method (SADABS software \cite{Sheldrick}) based on crystal shape (face indices) yielding very good agreement between equivalent reflections R$_{int}$ (see Table~\ref{tabl1}). The structure refinement was performed using SHELXL software \cite{Sheldrick} and included atomic coordinates, anisotropic atomic displacement parameters and  Ca:R occupation factors, assuming fully occupied sites and using 9 parameters in total.

Fig.~\ref{xray1} presents the evolution of lattice constants with rare earth concentration determined from single-crystal x-ray refinements performed at a fixed temperature of 250~K. Data points determined by neutron scattering experiments are also included. The scatter in both data sets arises due to uncertainties in measured WDS concentration $x$ ($\sim \pm 1\%$), measurement temperature ($250 \pm 5$~K) and systematic variations in measurement platforms (\ie, x-ray vs. neutron scattering), but is dominated by the uncertainty in temperature due to the extremely large thermal expansion (see below).

At 250~K with increasing $x$, the $a$-axis lattice parameter for all
$R$ increases at close to the same rate, while the variation of the $c$-axis shows a strong dependence on type of rare earth substituent. As shown in  Fig.~\ref{xray1}b), the $c$-axis remains constant with $x$ for La substitutions, but decreases for Ce, Pr and Nd at a rate that increases for smaller/heavier rare earth species. 
Overall, it is clear that the progression of the $c$-axis lattice parameter with rare earth substitution presents a unique opportunity to controllably apply chemical pressure by choice of rare earth substituent. Below, we investigate the effects of controlled $c$-axis reduction via this technique on the crystal structure and its instability to collapse.

%%%%%%%%%% FIGURE:  neutron raw data
\begin{figure}
\centering \resizebox{8cm}{!}{
  \includegraphics[width=16cm]{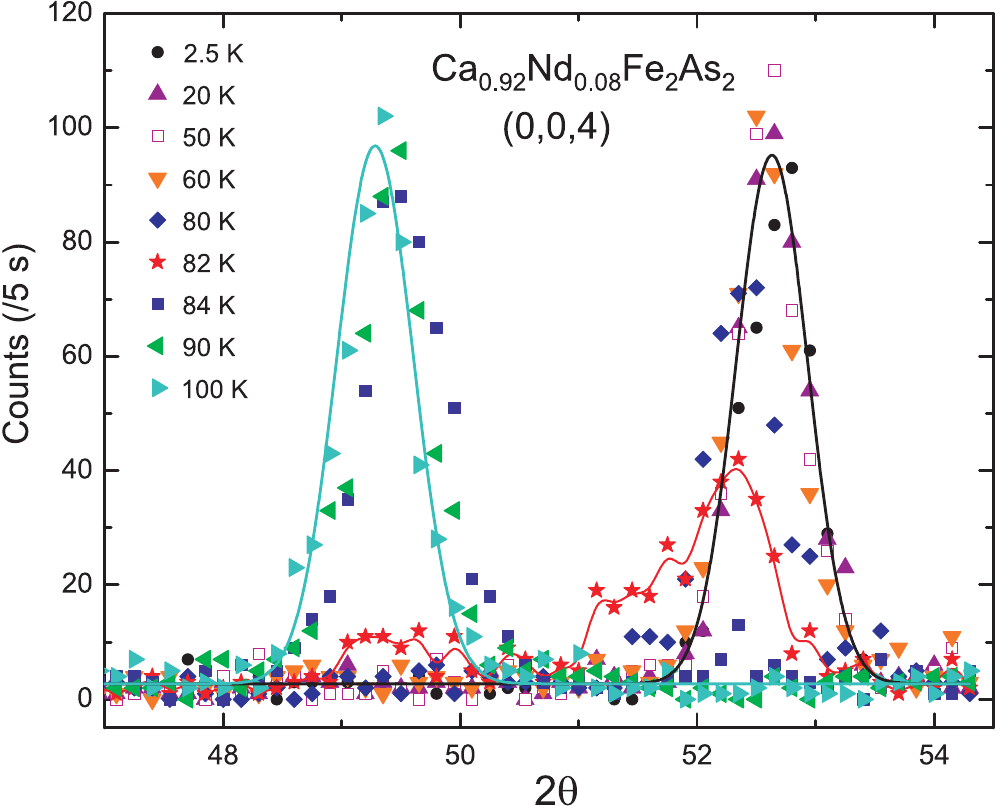}}
\caption{Neutron diffraction data for Ca$_{0.92}$Nd$_{0.08}$Fe$_2$As$_2$ summarized by
 a few selected\ $\protect\theta:2\protect\theta $ scans across the (004) structural Bragg reflection, showing the abrupt transition from the
collapsed to the uncollapsed tetragonal phase near 82~K. Data were obtained upon warming.}
\label{neutron1}
\end{figure}

%**Neutron scattering
Neutron diffraction experiments were performed on single-crystal samples using the BT-7 and BT-9 triple axis spectrometers at the NIST Center for Neutron Research. The incident energy was 14.7~meV using pyrolytic graphite (002) monochromators and analyzers. \ Data were collected using $\theta :2\theta $ scans to determine the temperature dependence of the $a$ and $c$ lattice parameters, in steps of 2 K. Typically data were taken upon warming and cooling through the range 2~K to 300~K in order to properly capture the large hysteresis of the structural collapse transition.

%%%%%%%%%%%%%%%%%%%%%%% Figure - NEUTRON CONTOUR
\begin{figure}[!t] \centering
  \resizebox{8cm}{!}{
  \includegraphics[width=8cm]{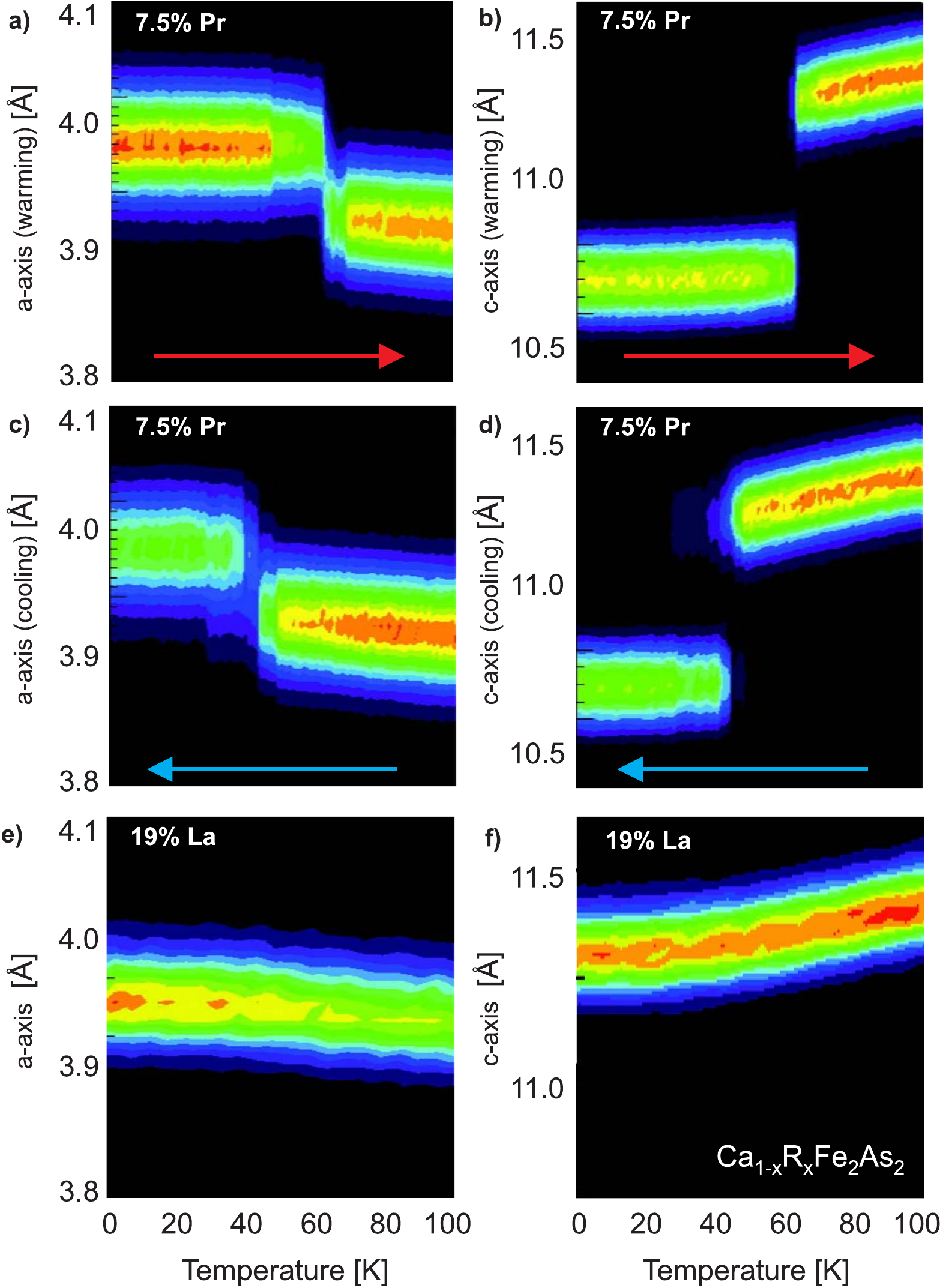}}
  \caption{\label{contour}  Effect of structural collapse of the tetragonal unit cell of \CaR\ by rare earth substitution on lattice parameters, determined by neutron diffraction measurements of (110) and (006) nuclear reflections are represented as $a$- and $c$-axis data, respectively. The color-intensity plots highlight the structural collapse of the tetragonal unit cell induced by rare earth substitution upon both warming (red arrows) and cooling (blue arrows) that is present in Ca$_{0.925}$Pr$_{0.075}$Fe$_2$As$_2$ (a-d), but absent in Ca$_{0.81}$La$_{0.19}$Fe$_2$As$_2$ (e-f). Data collected upon warming are displayed in panels a) and b) and for cooling in panels c) and d), emphasizing the hysteretic nature of the structural transition in the 7.5\% Pr crystal.}
\end{figure}

A few select neutron diffraction scans are shown in Fig.~\ref{neutron1} for a Nd$_{0.08}$Ca$_{0.92}$Fe$_{2}$As$_{2}$ crystal with mass 3~mg, measured upon warming. At base temperature there is a resolution limited peak on the high angle side, which has a modest shift to the left on warming as expected from thermal expansion. On warming, there is a dramatic structural transition that occurs between 80~K and 84~K. In particular, at 82~K there is a distribution of $c$-axis lattice parameters, while at 84~K the peak has jumped to smaller angle, indicating that the $c$-axis has suddenly increased. Above 84~K the system undergoes continuous thermal expansion as discussed below.

%%%%%%%%%%%%%%%%%%%%%%% Figure - C-AXIS DATA
\begin{figure}[!t] \centering
  \resizebox{8cm}{!}{
  \includegraphics[width=8cm]{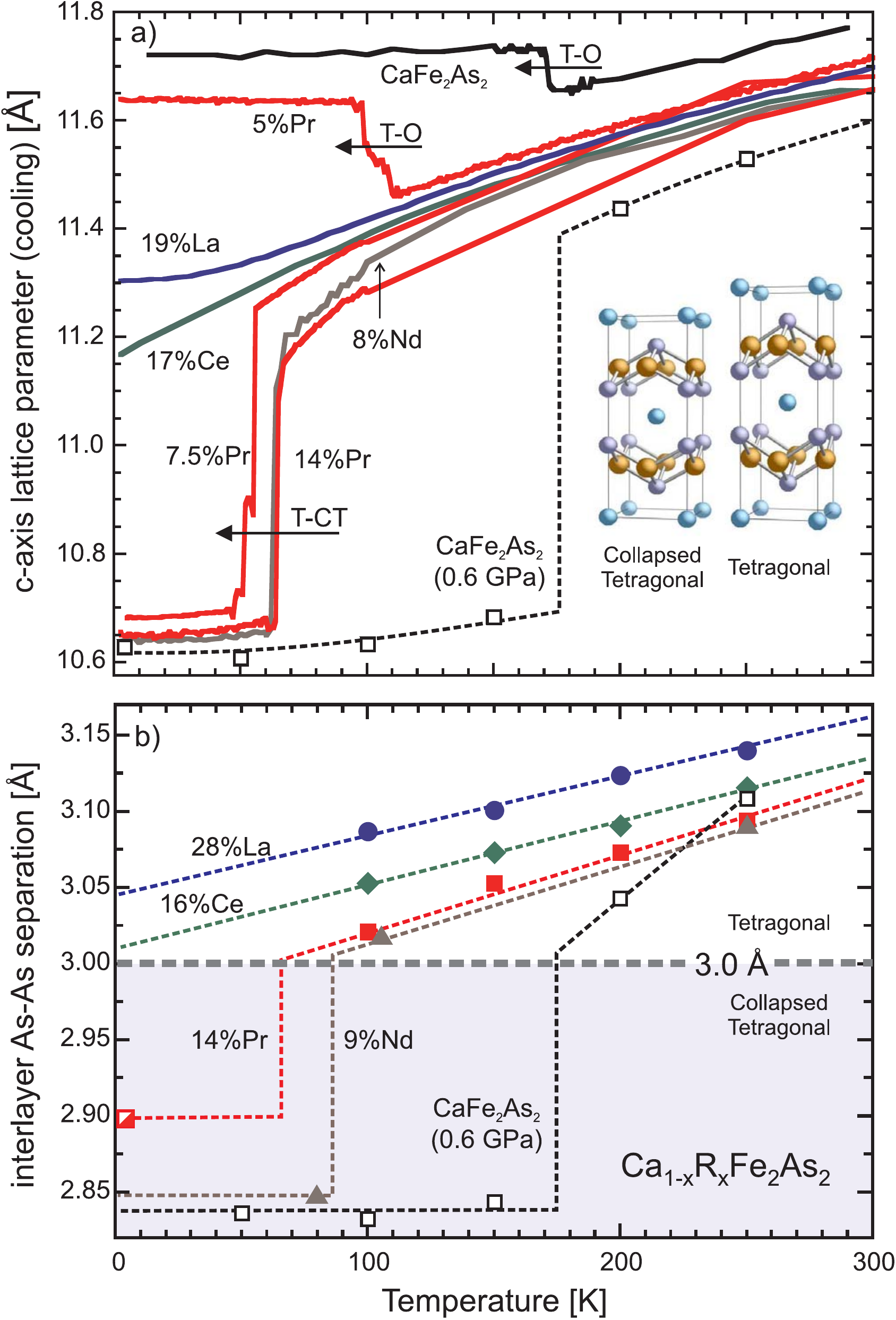}}
  \caption{\label{caxis} (a) $c$-axis lattice parameters of \CaR\ with R=La (blue), Ce (green), Pr (red) and Nd (brown), determined from neutron diffraction measurements of (006) nuclear reflections upon cooling. The evolution of the tetragonal-orthorhombic (T-O) and tetragonal-collapsed tetragonal (T-CT) structural transitions is compared to undoped \Ca\ at ambient pressure \cite{Kumar_Ni} (black line) and undoped \Ca\ under 0.6~GPa of applied hydrostatic pressure (open black squares; from Ref.~\onlinecite{Kreyssig}).
  (b) Comparison of the interlayer As-As separation distance for La, Ce, Pr and Nd-doped samples measured by single-crystal x-ray diffraction (filled symbols; open diagonal-square from powder neutron diffraction) and \Ca\ at 0.6~GPa (Ref.~\onlinecite{Kreyssig}). Dashed lines are guides, with vertical portions indicating the measured temperature of the CT transition.} 
\end{figure}

%** collapse vs T - contour plots
This abrupt shift in the (004) Bragg reflection arises from a dramatic shift in the $c$-axis lattice constant as a function of temperature, resulting from only 8\% substitution of Nd into \Ca. As shown in Fig.~\ref{contour}, the substitution of a similar amount of Pr into \Ca\ is also enough to drive the collapsed tetragonal (CT) transition, with the $a$-axis and $c$-axis lattice parameters undergoing a discontinuous jump that is hysteretic in temperature.  In contrast, the substitution of up to 28\% La does not drive the system toward any observable transition, consistent with the expectation that the larger ionic radius of La is not amenable to inducing positive chemical pressure.
The temperature dependence of the $c$-axis unit cell dimension upon cooling for a series of R-doped crystals, shown in Fig.~\ref{caxis}a), presents a summary of this dramatic variation in structural properties. As shown, small amounts of Nd and Pr substitutions act in a quantitatively similar manner, achieving an almost identical collapse transition, while crystals with larger rare earth size substitution (\ie, 19\% La and 17\% Ce) fail to collapse through the entire temperature range studied. 
Note that we have verified that the CT transition is intrinsic to these crystals, and not caused by strain fields induced by growth conditions as observed in undoped \Ca.\cite{Ran-Canfield} (see Fig.~\ref{flux} in the next section for more details).

%%%%%%%%%%%%%%%%%%%%%%% FIGURE:  22%Ce He-gas neutron experiment
\begin{figure} \centering
  \resizebox{8cm}{!}{
  \includegraphics[width=8cm]{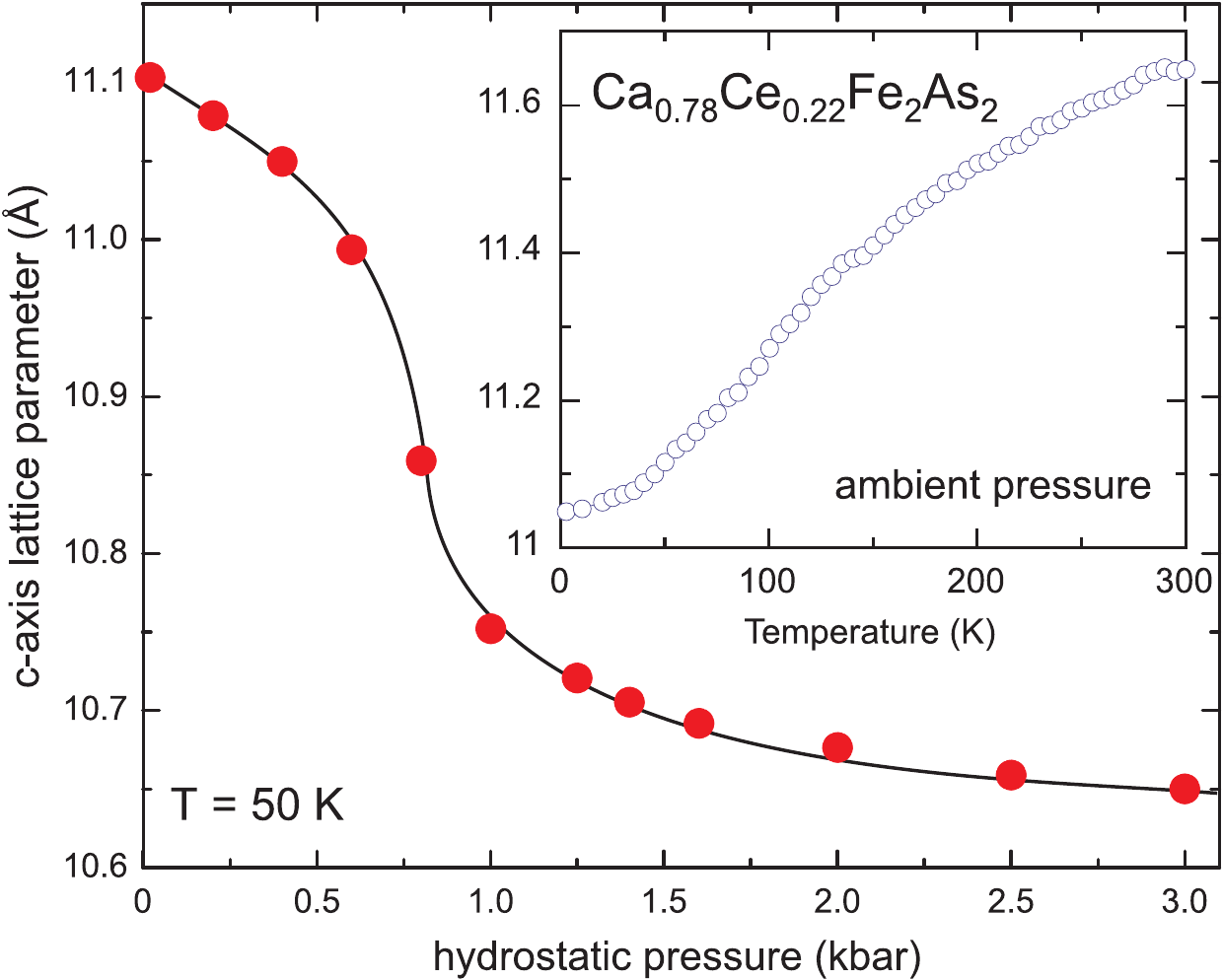}}
  \caption{\label{Ce_pressure} Pressure-induced collapse of the $c$-axis lattice parameter in a Ca$_{0.78}$Ce$_{0.22}$Fe$_2$As$_2$ crystal, as obtained by neutron diffraction measurements of the sample under an applied He-gas hydrostatic pressure at a constant temperature of 50~K. The abrupt change above 0.5~kbar is the collapse of the tetragonal unit cell, as discussed in the main text. Inset: ambient pressure $c$-axis measurement of the same sample in the non-collapsed state, indicating an unprecedented 5.3\% thermal expansion (see text).}
\end{figure}

The absolute change in $c$-axis upon collapse is nearly identical to that observed in \Ca\ under pressure \cite{Kreyssig}, despite subtle but important differences in charge doping effects.  This speaks to the dominant bonding interactions driving the collapse, also apparent in the extremely large $c$-axis thermal expansion observed in this series: even in the absence of a collapse transition, a 22\% Ce-doped crystal undergoes a 5.3\% expansion of the $c$-axis between 0 and 300~K, giving a linear thermal expansion coefficient of 180$\times$10$^{-6}$/K. This value is one of the largest for any metal (\eg, as compared to the largest known thermal expansion values of 97, 83 and 71$\times$10$^{-6}$/K for elemental Cs, K and Na, respectively, at 25$^o$C),\cite{CRC} and even rivals the largest known values for any solid as observed in molecular crystals.\cite{Das}

%%%%%%%%%% FIGURE:  xray2 - internal structure
\begin{figure*}[!t] \centering
  \resizebox{16cm}{!}{
  \includegraphics[width=16cm]{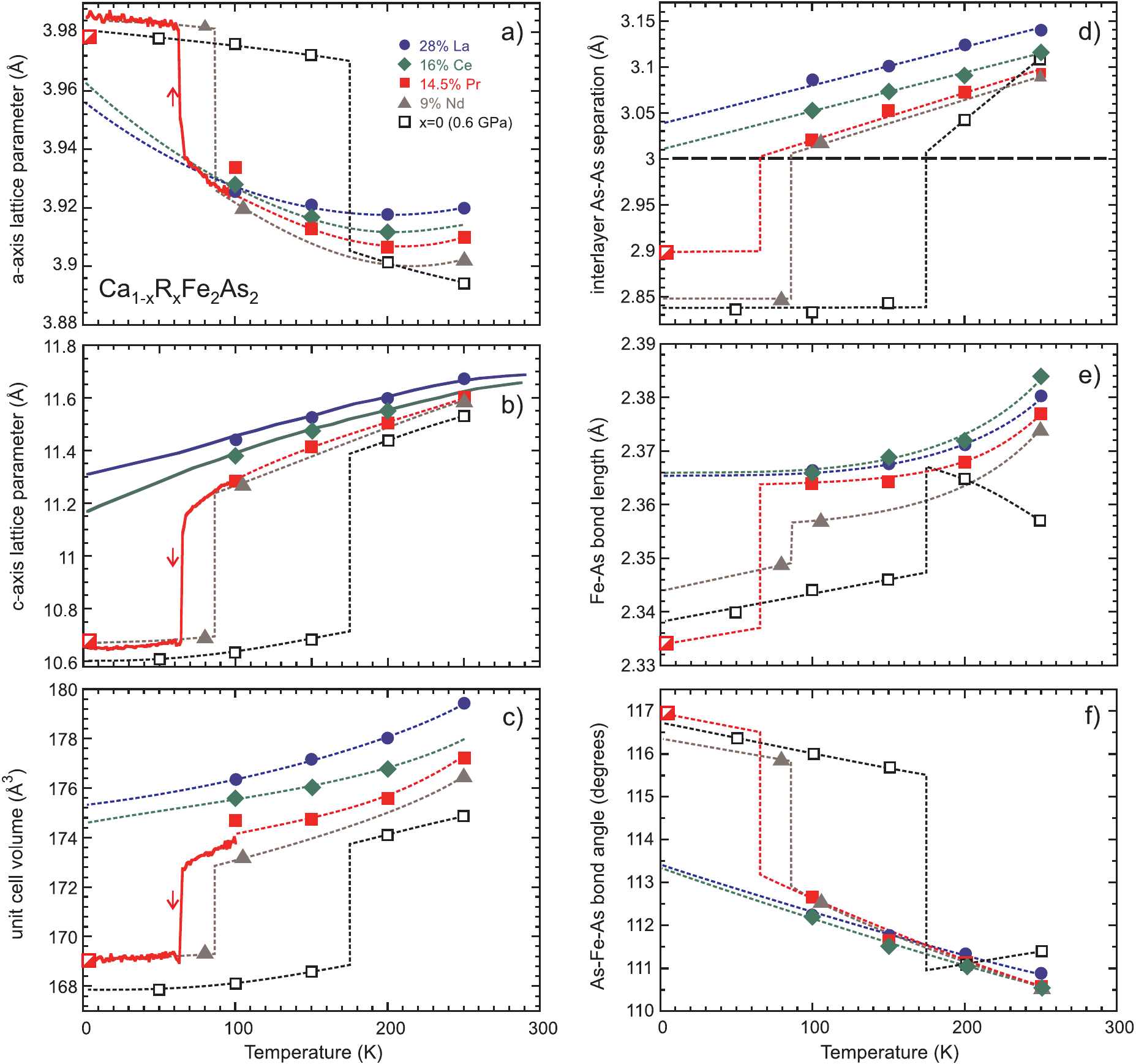}}
  \caption{\label{xray2} Structural characterization of unit cell and substructure of \CaR\ for several characteristic crystals, including single-crystal neutron data (solid lines) extracted from Bragg reflections (see text), refinement of neutron powder diffraction for 14.5\% Pr (red open diagonal-square), as well as refinement data from single-crystal diffraction (solid symbols). Dashed lines are guides to single-crystal x-ray refinement results, with vertical portions indicating the temperature of the structural collapse transition determined from magnetic susceptibility and arrows indicating cooling direction. Open squares are data for \Ca\ under hydrostatic pressure reproduced from Ref.~\onlinecite{Kreyssig} for comparison.}
\end{figure*}

The strong $c$-axis contraction and instability to collapse is driven by an increasing overlap of interlayer As orbitals,\cite{Hoffman,Yildirim} controlled via the chemical pressure instilled by rare earth substitution. In fact, the As-As interlayer separation itself appears to be the key parameter controlling the collapse: as shown in Fig.~\ref{caxis}b), both \Ca\ under 0.6~GPa pressure as well as Pr- and Nd-doped \Ca\ crystals collapse once the interlayer As-As distance reaches a critical value of $\sim 3.0$ \AA, while La- and Ce-doped crystals which remain uncollapsed to the lowest temperatures do not cross this value.

%%%%%%%%%%%%%%%%%%%%%%% Table 1
\begin{table*}[tbh]

\caption{\label{tabl1}Crystallographic data for
Ca$_{0.91}$Nd$_{0.09}$Fe$_2$As$_2$ determined by single-crystal x-ray
diffraction at 250~K (tetragonal structure (T), above collapse transition), 105~K (T structure, just above collapse transition), and 80~K (collapsed tetragonal (CT) structure, just below collapse transition). Uncertainty in temperature values is $\pm 5$~K.}

\begin{ruledtabular}
\begin{tabular}{llll}
%\begin{tabular*}{\textwidth}{@{}l*{15}{@{\extracolsep{0pt plus12pt}}l}}
Temperature&80~K&105~K&250~K\\
\hline
Structure&CT&T&T\\
Space group&I4/mmm&I4/mmm&I4/mmm\\
$a$($\mathrm{\AA}$)&3.9822(16)&3.9202(15)&3.9025(7)\\
$b$($\mathrm{\AA}$)&=$a$&=$a$&=$a$\\
$c$($\mathrm{\AA}$)&10.684(4)&11.273(5)&11.591(2)\\
$V$($\mathrm{\AA}^3)$&169.43(6)&173.24(5)&176.53(5)\\
$Z$&1&1&1\\
Density(g/cm$^3$)&6.096 &5.964&5.865\\
Refl.collected&927&1128&1639\\
Independent refl.&96&99&100\\
R$_{int}$\footnotemark[1] (\%)&3.38&3.69&3.93\\
wR$_2$\footnotemark[2], all refl.&2.25&5.74&6.61\\
R$_1$\footnotemark[3], I$\geq 2\sigma$I&5.11&2.58&3.15\\
Atomic parameters:&&&\\
Nd occupation factor&0.086(7)&0.091(7)&0.097(8)\\
Ca/Nd&2$a$(0,0,0)&2$a$(0,0,0)&2$a$(0,0,0)\\
Fe&4d(0,1/2,1/4)&4d(0,1/2,1/4)&4d(0,1/2,1/4)\\
As&4e(1/2,1/2,$z$)&4e(1/2,1/2,$z$)&4e(1/2,1/2,$z$)\\
&$z$=0.13328(11)&$z$=0.13391(12)&$z$=0.13339(14)\\
Atomic displacement&&\\
parameters U$_{eq}$ ($\mathrm{\AA}^2$):&&&\\
Ca1/Nd1&0.0089(8)&0.0110(9)&0.0173(10)\\
Fe1&0.0079(3)&0.0097(4)&0.0158(5)\\
As1&0.0074(3)&0.0094(3)&0.0158(4)\\
Bond lengths ($\mathrm{\AA}$):&&\\
Ca/Nd-As&3.1554(12)$\times$8&3.1563(12)$\times$8&3.1631(9)$\times$8\\
Fe-As&2.3494(10)$\times$4&2.33568(7)$\times$4&2.3736(10)$\times$4\\
Fe-Fe&2.8158(11) &2.7720(11) &2.7595(5) \\
Bond angles (deg):&&&\\
As-Fe-As&115.88(6)$\times$2&112.54(6)$\times$2&110.58(7)$\times$2\\
 &106.37(3)$\times$4&107.96(3)$\times$4&108.92(3)$\times$4\\
Fe-As-Fe&73.63(3)$\times$4&72.04(3)$\times$4&71.08(3)$\times$4\\
%\mr
\hline
\end{tabular}
\footnotesize\rm
\footnotetext[1]{~R$_{int}$= $\Sigma \vline F_o^2$-$F_c^2$(mean)/$\Sigma$[$F_o^2$]}
\footnotetext[2]{~wR$_2$ = {$\Sigma $[w($F_o^2$ -$F_c^2$)$^2$] / $\Sigma $[w($F_o^2$)$^2$]}$^{1/2}$}
\footnotetext[3]{~R$_1$ = $\Sigma \parallel F_o \vline $ - $\vline F_c\parallel$ / $\Sigma \vline F_o \vline$} 
\end{ruledtabular}
\end{table*}

%% neutron pressure exp't
The highest-doped Ce compound is just on the verge of collapse at ambient pressure.
Applying a tiny amount of external pressure to a 22\% Ce crystal, whose As-As separation approaches $3.0$ \AA\ at zero temperature, confirms this scenario.The $c$-axis lattice parameter of a 22\% Ce sample was studied via neutron diffraction as a function of applied hydrostatic pressure achieved using an Al-alloy He-gas pressure cell as described previously.\cite{Goldman} As shown in Fig.~\ref{Ce_pressure}, a collapse is induced at only 0.05~GPa applied pressure at 50~K constant temperature. This is an order of magnitude lower pressure than that required to induce the collapse in undoped \Ca, \cite{Goldman} which is easily understood by the fact that 22\% Ce substitution brings the crystal structure very close to the critical $3~\AA$ interlayer As-As distance at low temperatures, thereby only requiring a very small additional pressure to induce the collapse.

Furthermore, it appears the critical distance has more to do with the $p$-orbital bonding character than the exact chemical make-up: phosphorus-based materials SrRh$_2$P$_2$ and EuRh$_2$P$_2$ both undergo a collapse of the $c$-axis dimension by $\sim 1.5$~\AA\ as a function of pressure (5~GPa) and temperature (800~K), respectively, when they cross a similar critical interlayer P-P distance of $\sim$3~\AA.\cite{Huhnt} The fact that 3~\AA\ is the average value between covalent and Van der Waals radii of both elemental As and P (Ref.~\onlinecite{CRC}) suggests that this striking universal behavior can be observed in any system with similar $p$-orbital overlap approaching this critical value.

%%% TRANSPORT/SUSCEPTIBILITY -4 panel
\begin{figure*}[!t] \centering
  \resizebox{16cm}{!}{
  \includegraphics[width=16cm]{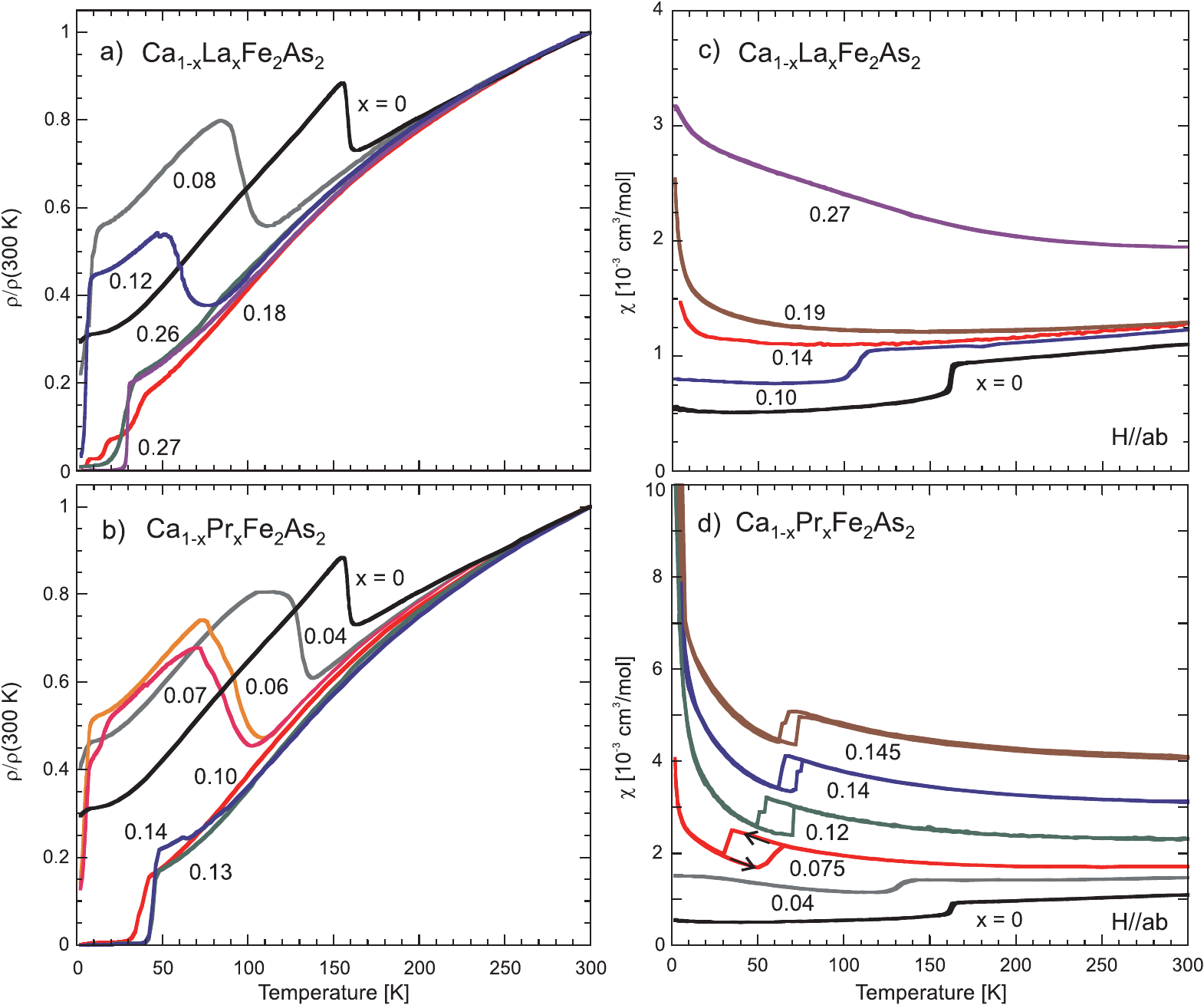}}
  \caption{\label{ressusc} 
  The normalized electrical resistivity of \CaR\ crystals with various concentrations of La (a) and Pr (b) tracks the evolution of magnetic, structural and superconducting phase transitions. The tetragonal/paramagnetic to orthorhombic/antiferromagnetic transition in \Ca, indicated by a sharp rise in $\rho(T)$ near 165~K (black curves), is suppressed with R substitution before entering a superconducting phase at higher $x$. Pr substitution also induces a collapse of the tetragonal structure at $x>0.07$, barely seen in the $x=0.14$ sample as a kink in $\rho(T)$.
  Magnetic susceptibility data for La (c) and Pr (d) substitutions (measured in fields $\geq 0.1$~T for Pr and $\geq 1$~T for La) exhibits sharp features associated with the AFM transition (drop in $\chi(T)$) in both cases, in addition to the collapsed tetragonal transition (hysteretic drops in $\chi(T)$) for Pr substitutions. Data for Pr-doped samples $x$= 0.12, 0.14 and 0.145 are shifted up for clarity. }
\end{figure*}

%% xray table
Full structural refinement data for 9\% Nd at temperatures above and below the collapse transition are tabulated in Table~\ref{tabl1} and presented graphically in Fig.~\ref{xray2} along with several other characteristic samples for various temperatures and rare earth concentrations.  For each characteristic doping, the same crystal was measured at several fixed temperatures, providing a more systematic set of data that suffer from less scatter than the $x$-dependent quantities shown in Fig.~\ref{xray1}. Data are presented for each rare earth doping and compared to available data for undoped \Ca\ under 0.6~GPa applied pressure.\cite{Kreyssig} The intralayer Fe-As bond length, which remains relatively rigid in all FeAs-based compounds, is shown to decrease with decreasing temperature in 28\% La, 16\% Ce, 14.5\% Pr, and 9\% Nd from 250~K down to 100~K, as shown in Fig.~\ref{xray2}(e). Samples that undergo a structural collapse show a large contraction of the Fe-As bond length in line with the concomitant expansion of the $a$-axis plane.
Interestingly, as shown in Fig.~\ref{xray2}(f) the As-Fe-As tetrahedral bond angle shows an even stronger evolution with temperature in all samples. While this angle also shows an abrupt increase through the collapse transition in Pr- and Nd-doped samples as expected by the strong $c$-axis contraction, even the non-collapsing samples show a $\sim 2\%$ increase from 250~K to zero temperature indicative of the strong thermal expansion discussed above.

%%%%%%%%%%%%%%%%%%%%%%%%%%%%%%%%%%%%%%%%%%%%%%%%%%%%%%%%%%%%%%%%%%%%%%%%%%%%%%%%%%
\section{Transport and Susceptibility}

In addition to structural tuning, rare earth substitution introduces an important extra degree of freedom that applied pressure does not: charge doping via aliovalent substitution allows for fine-tuning of both pressure and doping effects on the physical properties. Here we use electrical resistivity $\rho$ measured with the standard four-probe ac method, and magnetic susceptibility $\chi$ measured in a SQUID magnetometer, to track the evolution of both structural and electronic properties as a function of rare earth substitution.

%RESISTIVITY
Fig.~\ref{ressusc}(a) presents the progression of electrical resistivity $\rho$ of single crystals of \LaCa\ for various La concentrations, normalized to 300~K values. 
In \Ca, the sharp jump at $T_N=165$~K is due to a structural phase transition from tetragonal to orthorhombic upon cooling, and is known to coincide with the onset of antiferromagnetic (AFM) order.\cite{Ronning,Ni78,Goldman78} The substitution of La into \CaR\ suppresses the feature associated with $T_N$ to lower temperatures, eventually suppressing it completely with higher La concentrations.  As shown in Fig.~\ref{ressusc}(b), Pr substitution also acts to suppress $T_N$ to lower temperatures in a similar manner.

%%%%%%%%%% FIGURE:  compare-FeAs-Sn-Flux
\begin{figure} \centering
  \resizebox{8cm}{!}{
  \includegraphics[width=8cm]{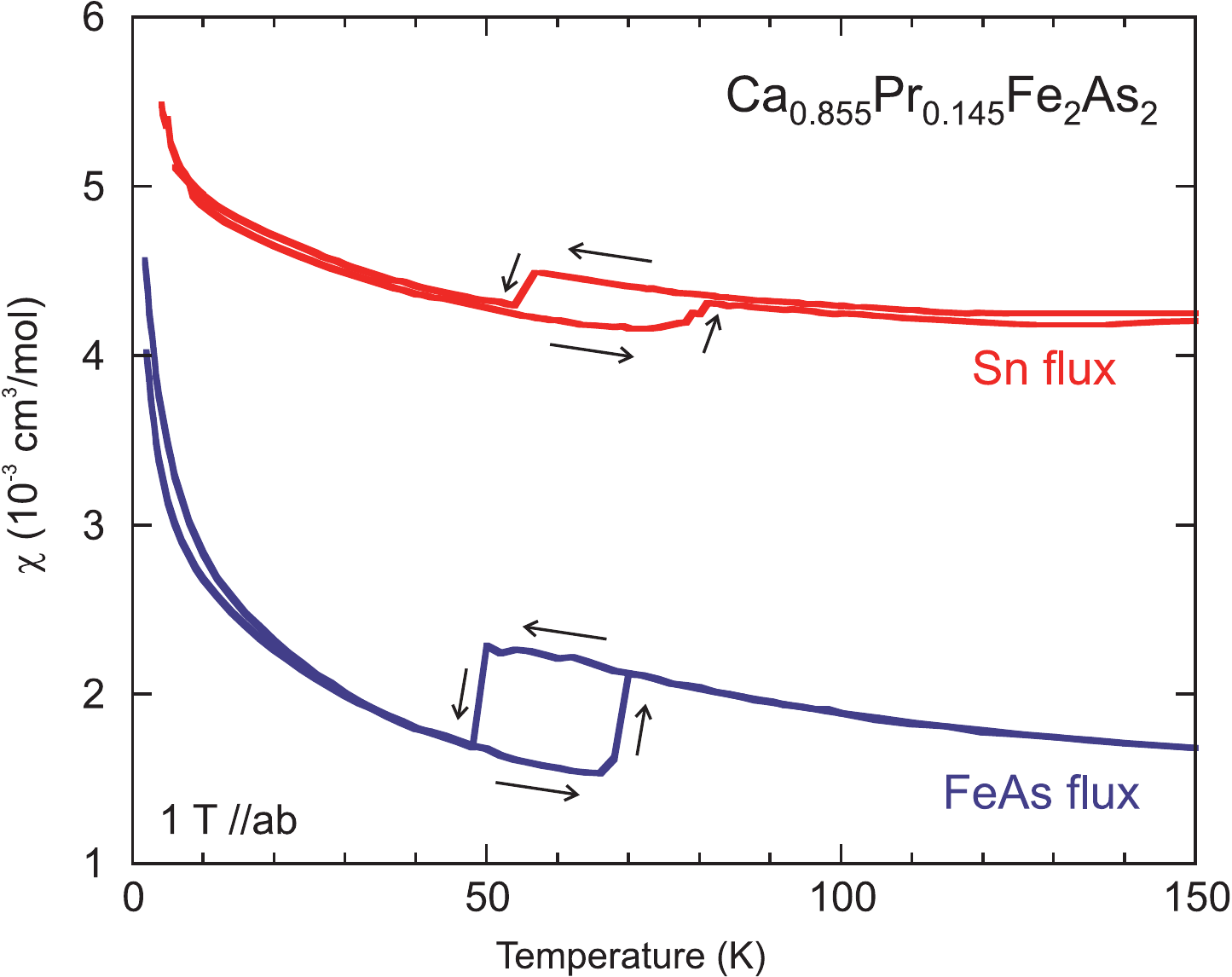}}
  \caption{\label{flux} Comparison of structural collapse transition between crystals of Ca$_{0.855}$Pr$_{0.145}$Fe$_2$As$_2$ grown using FeAs flux (blue) and Sn flux (red). Arrows indicate warming/cooling directions.}
\end{figure}

%SUSCEPTIBILITY
This trend is also observed in magnetic susceptibility data, shown in Figs.~\ref{ressusc}(c) and (d). Closest to the critical La concentration for suppression of the AFM phase, such as at $x=0.14$, $\chi(T)$ shows an extremely flat temperature dependence that mimics the undoped compound above $T_N$. A moderate increase in $\chi(T)$ is apparent at low temperatures, and is attributed to the FeAs sublattice of the crystals given the non-magnetic nature of La substituents. In contrast, $\chi(T)$ enhances strongly with Pr substitution due to the increasing concentration of Pr localized 4$f$ electrons and their contribution of a Curie-like susceptibility, as shown in Fig.~\ref{ressusc}(d). Pr substitution gradually suppresses the step-like feature at $T_N$ as with La, but an abrupt appearance of another first-order transition occurs near 7.5\%, coinciding with the structural CT transition.

%Sn- vs self-flux
A comparison of crystals grown under different conditions was performed to verify that strain mechanisms are not the primary cause of the CT transition. This phenomenon was recently reported to occur under different thermal treatments in pure \Ca, resulting in replication of conditions similar to applied pressure that stabilize the structural collapse.\cite{Ran-Canfield} Fig.~\ref{flux} presents a comparison of susceptibility data for two crystals with the same Pr concentration but grown under different conditions: one using the FeAs (self) flux technique described above and another using Sn flux, known to provide the least amount of strain during growth and cooling.\cite{Ran-Canfield} As shown in the $\chi(T)$ curves, the structural collapse is evident in both cases, and moreover is almost identical, thus indicating that the collapse is caused by intrinsic chemical pressure due to rare earth substitution and not due to an extrinsic strain field as found for undoped \Ca.

\begin{figure}[!t] \centering
  \resizebox{8cm}{!}{
  \includegraphics[width=8cm]{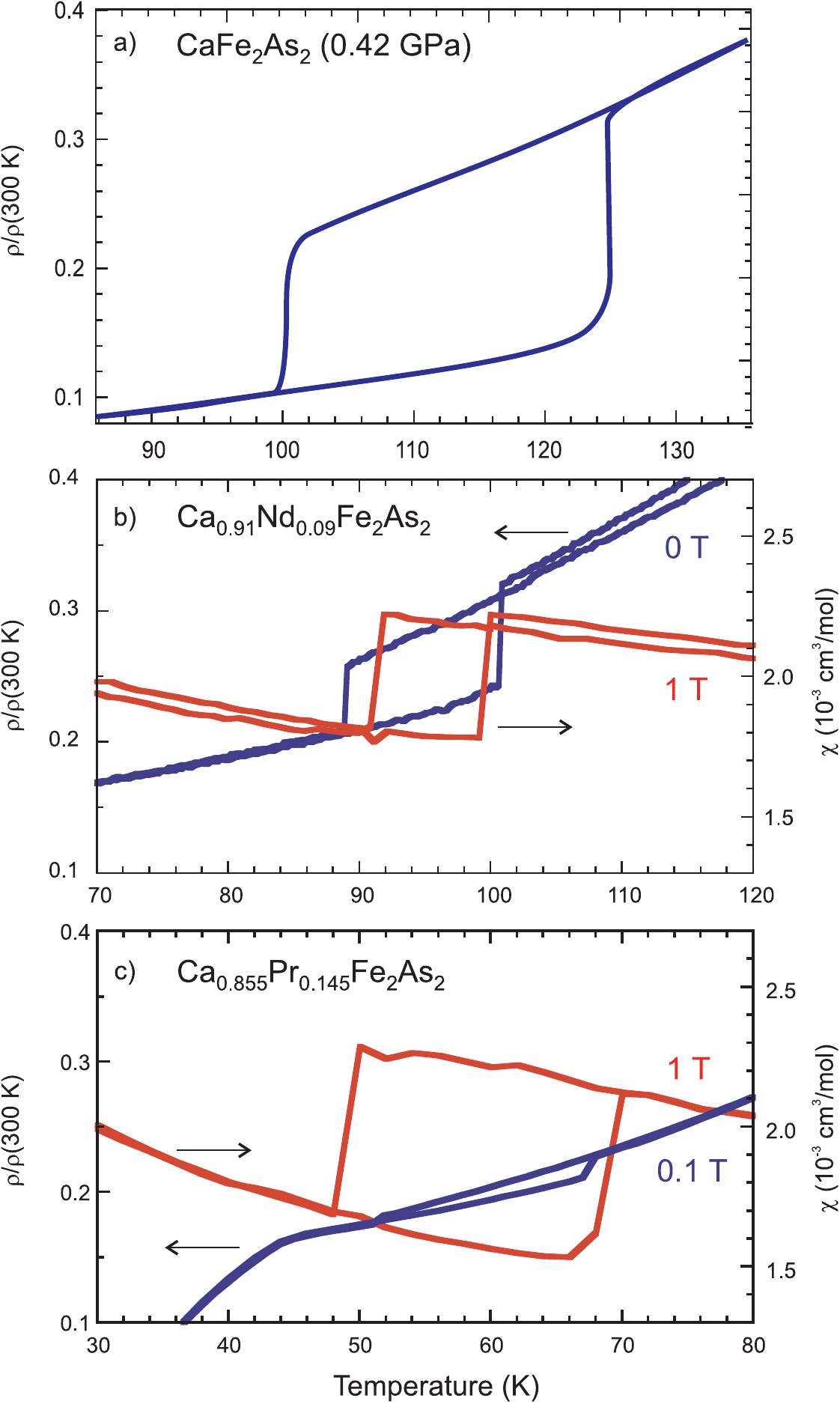}}
  \caption{\label{collapse} Comparison of transport and susceptibility data on warming and cooling through the structural collapse transition in three representative samples: a) resistivity of undoped \Ca\ under 0.42~GPa applied hydrostatic pressure (reproduced from Ref.~\onlinecite{Yu}); b) Ca$_{0.91}$Nd$_{0.09}$Fe$_2$As$_2$ at ambient pressure; and c)  
  Ca$_{0.855}$Pr$_{0.145}$Fe$_2$As$_2$ at ambient pressure. Note the normalized resistivity (left) and susceptibility (right) vertical scales are equal for each panel to allow comparison of relative change of $\rho(T)$ through the collapse transition.}
\end{figure}

A large hysteresis of $\sim$30~K between temperature cooling and warming measurements is present in all structural, transport and magnetic measurements, indicating the first-order nature of this transition. The magnetic character of this hysteresis does not appear to depend on the rare earth species (c.f. only present in Pr- and Nd-doped crystals), as indicated by the abrupt, sharp transitions present in $\chi(T)$ as highlighted in Fig.~\ref{collapse}. However, the change in absolute resistivity appears to show a progression from a very large increase(decrease) on warming(cooling) in undoped \Ca\ (under applied pressures) to an almost negligible change in magnitude at large rare earth concentration. As shown in Fig.~\ref{collapse}, a change in $\rho(T)$ of the order of 10\% of the normalized resistivity is observed in undoped \Ca\ under pressure,\cite{Yu} as compared to a much smaller change in a crystal with 9\% Nd and an almost negligible change in magnitude as shown for 14.5\% Pr. This appears to be related to the effect of electron doping caused by trivalent rare earth substitution, as the magnitude of the change decreases with increasing rare earth substitution.

%%%%%%%%%%%%%%%%%%%%%%% FIGURE:  collapse zoom

\begin{figure}[!t] \centering
  \resizebox{8cm}{!}{
  \includegraphics[width=8cm]{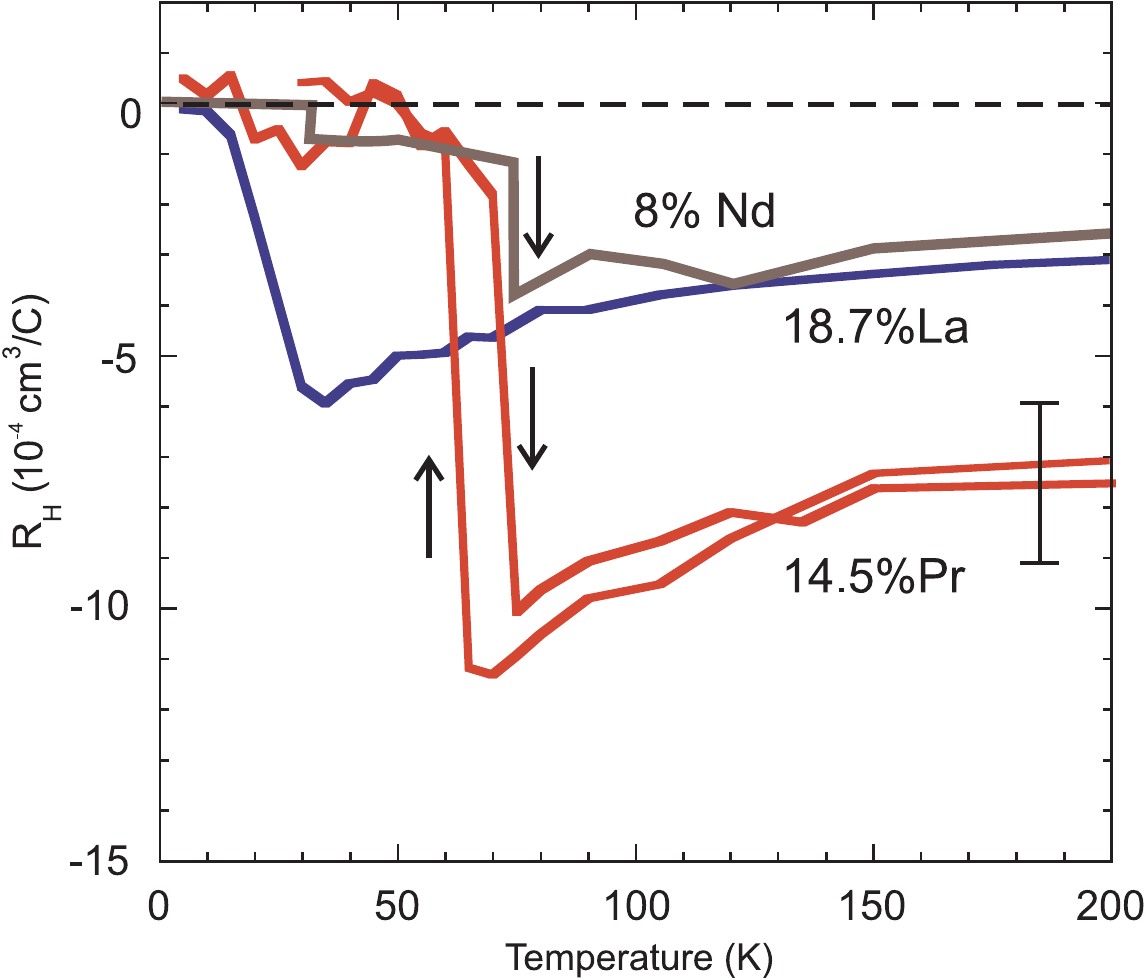}}
  \caption{\label{hall} Temperature dependence of Hall effect in 14.5\% Pr and 8\% Nd crystals through their collapse transitions, showing a dramatic change in the Hall coefficient through the collapse, compared to a 19\% La crystal that does not undergo a collapse. Black error bar denotes the resolution of the experiment, and arrows indicate direction of warming/cooling during experiment.}
\end{figure}

%**** HALL DATA
Through the structural collapse, a dramatic change in electronic structure is predicted to occur \cite{Hoffman,Yildirim,Goldman} involving the elimination of a cylindrical hole pocket centered at the $\Gamma$ point in the Brillouin zone. 
This is confirmed by measurements of the Hall coefficient $R_H$ in 14.5\% Pr shown in Fig.~\ref{hall}, which provide evidence for a dramatic and abrupt change in electronic structure through the collapse transition, even exhibiting hysteretic behavior identical to that observed in structural (x-ray and neutron scattering) and magnetic (susceptibility) data. The reduction of $R_H$ towards zero below the CT transition suggests a transformation toward an almost exact compensation of electron and hole bands. In contrast, there is very little relative change in longitudinal resistivity through the structural transition as noted above. For example, note the contrast in behavior for 14.5\% Pr, which shows a dramatic order of magnitude drop in $R_H(T)$ through the CT transition as shown in Fig.~\ref{hall}, while there is an almost negligible relative change in $\rho(T)$ as shown in Fig.~\ref{collapse}(c). This suggests that the change in band structure through the collapse does not dramatically affect the bands that dominate intra-layer $ab$-plane transport, which is consistent with an iron plane that expands but nevertheless remains intact through the collapse with a lower electronic density of states.\cite{Yildirim,Goldman} Moreover, because electron doping decreases the amplitude of the jump in $\rho(T)$ at the CT transition, which is most pronounced in undoped \Ca\ under pressure \cite{Yu} and almost absent in the 14\% Pr crystal (c.f. Fig.~\ref{collapse}), one can conclude that the enlarging electron bands are more two-dimensional than the shrinking hole band(s), leading to this effect. More work is required to fully explore the change in the electronic structure through the CT transition, but ambient-pressure access to both phases as a function of a continuously tunable parameter such as temperature promises to provide much insight into the nature of the bonding that dominates the electronic and magnetic properties of these materials.

%%%%%%%%%%%%%%%%%%%%%%% Figure - HALL EFFECT

%%%%%%%%%%%%%%%%%%%%%%%%%%%%%%%%%%%%%%%%%%%%%%%%%%%%%%%%%%%%%%%%%%%%%%%%%%%%%%%%%%
\section{Superconducting Phase}

\begin{figure}[!t] \centering
  \resizebox{8cm}{!}{
  \includegraphics[width=8cm]{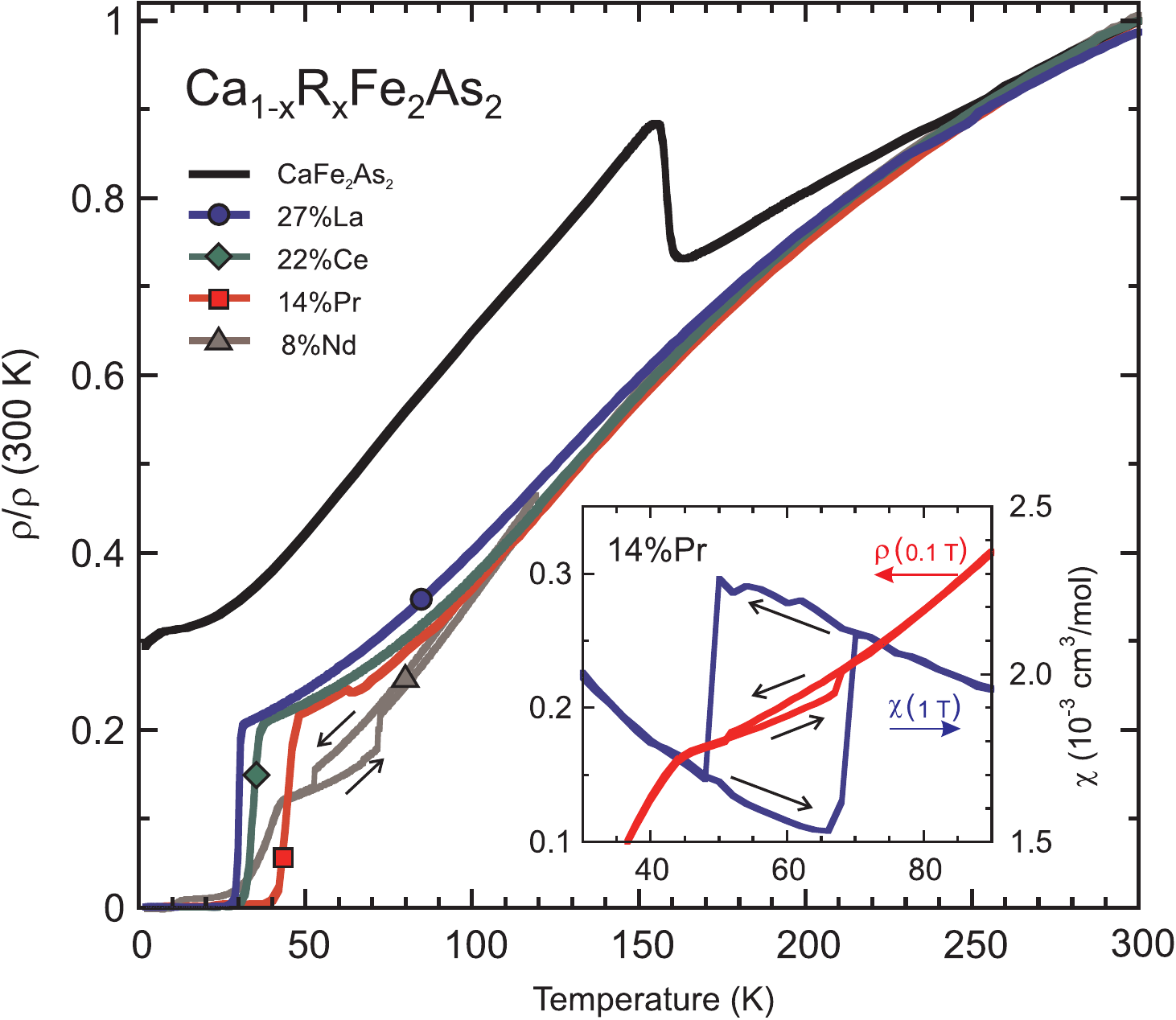}}
  \caption{\label{resistivity} Resistivity of representative samples of \CaR\ with La, Ce, Pr and Nd substitution, as compared to that of undoped \Ca. 
  The hysteretic structural transition, most clearly observed in the 8\% Nd crystal, is marked by arrows (warming-up, cooling-down). Inset: characterization of transitions in a 14\% Pr crystal, showing both hysteretic features from the structural collapse transition and the onset of superconductivity in resistivity near 45~K.}
\end{figure}

As evident in Fig.~\ref{ressusc}, the resistivity of Pr- and La-doped samples exhibits high-temperature superconductivity, with the highest observed onset temperature reaching 47~K as shown in Fig.~\ref{resistivity}. 
This is much higher than any value reported previously in electron-doped intermetallic FeAs-based systems, including both the commonly employed transition metal doping and the only other previously known case of electron doping via rare earth substitution in Sr$_{1-x}$La$_x$Fe$_2$As$_2$.\cite{Muraba} It also surpasses the highest values found in hole-doped (\eg, K$^{1+}$ for Ba$^{2+}$, Ref.~\onlinecite{Rotter}) 122 compounds which have a maximum \tc\ of $\sim 38$~K, and is more comparable to that found in the fluorine-based materials (Ca,R)FeAsF,\cite{Cheng} where similar rare earth electron-doping with Pr and Nd results in \tc\ values approaching the highest reported for any non-cuprate material.

The appearance of superconductivity in the \CaR\ series is consistent with the generally accepted hypothesis that the minimization of chemical disorder in the active FeAs layers, by substitution in the alkaline earth site, allows for the highest possible \tc\ values in the iron-based materials. What is most surprising, however, is that this superconducting phase exists in {\it both} collapsed and uncollapsed structures. As shown in Fig.~\ref{resistivity}, superconductivity is present both in 14\% Pr and 8\% Nd crystals which both undergo a collapse transition, and in 27\% La and 22\% Ce crystals which do not collapse. Given the evidence for a substantial change in the electronic structure through the structural collapse as shown by Hall data in Fig.~\ref{hall}, it is remarkable that high-temperature superconductivity appears to occur in this system regardless of the nature of the interlayer bonding: an insensitivity of pairing in the iron layer to the configuration of As $p$-orbitals would provide strong constraints on a microscopic model of superconductivity originating in the iron sublattice.
Furthermore, because theoretical calculations predict a non-magnetic ground state in the CT phase,\cite{Yildirim} it is tantalizing to conclude that superconductivity is originating in a phase void of spin fluctuations, providing an additional pivotal constraint on the nature of the pairing mechanism. However, the same calculation also predicts a second nearly degenerate magnetic ground state for the CT phase, so the perturbation introduced by charge doping must be properly included before such conclusions are made. 

Experimentally, the coexistence of a small fraction of non-collapsed tetragonal phase cannot be ruled out below the $\sim 1\%$ level from our elastic neutron scattering data, although it is highly unlikely due to the dramatic difference in lattice constants between the two structural phases. Nevertheless, the accessibility of the CT phase at ambient pressures in \CaR\ will allow for magnetism to be probed in a manner similar to that done for undoped \Ca\ under pressure.\cite{Goldman} Below, we investigate and rule out several possible extrinsic causes of superconductivity that would point to other explanations.

\begin{figure}[!t] \centering
  \resizebox{8cm}{!}{
  \includegraphics[width=8cm]{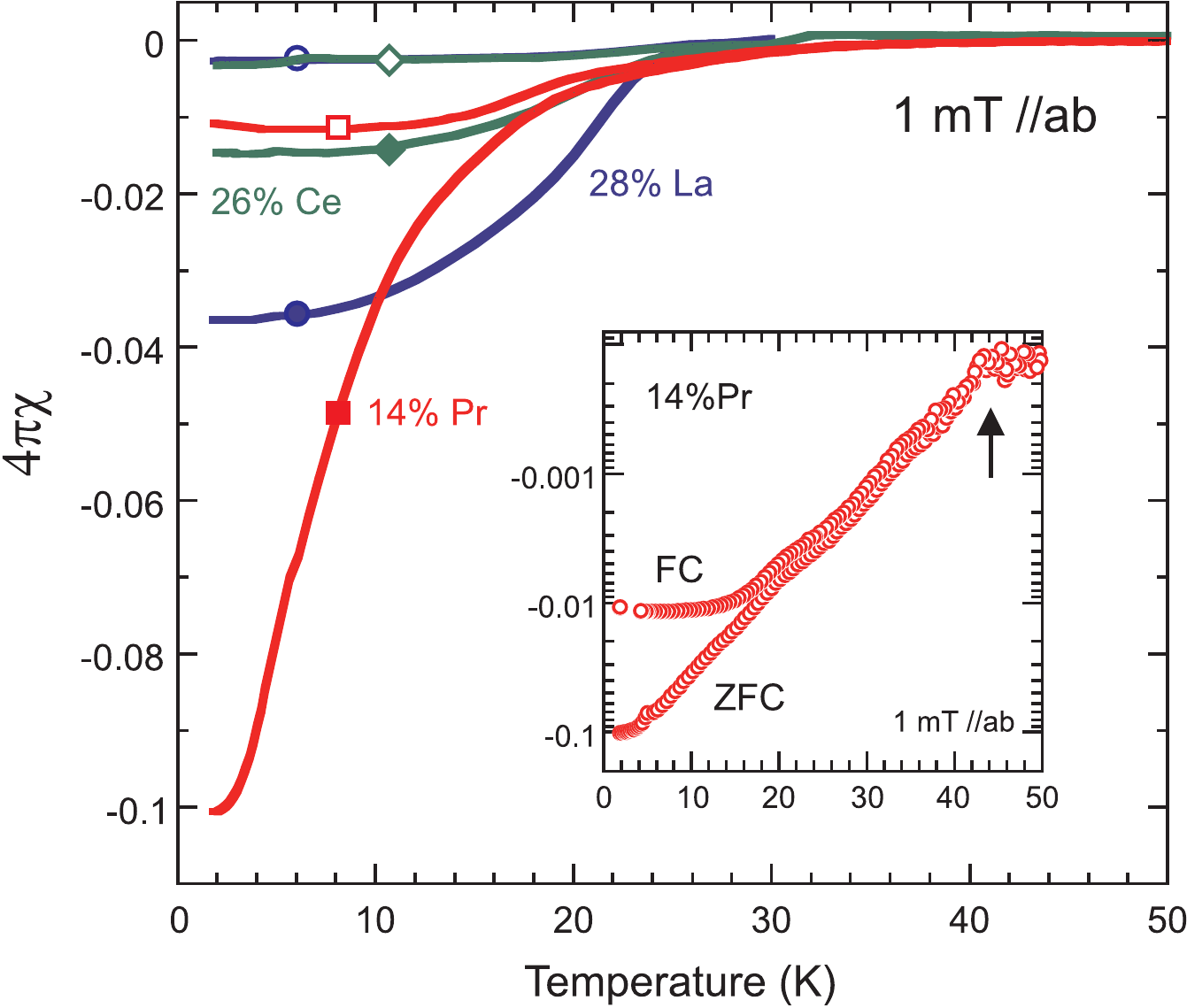}}
  \caption{\label{screening} Determination of superconducting volume fractions from diamagnetic screening estimation of three characteristic samples of \CaR, with La (blue-circles), Ce (green-diamonds), and Pr (red-squares). Closed symbols indicate zero-field-cooled data, and open symbols indicate field-cooled data. Inset: onset of diamagnetic screening in magnetic susceptibility of a 14\% Pr crystal at $T_c=44$~K (arrow), plotted on a semi-log scale to expose the onset temperature.}
\end{figure}

Superconducting transitions are observed in all rare earth substitutions as evidenced by resistive transitions and the onset of Meissner screening in magnetic susceptibility. 
As shown in Fig.~\ref{screening}, three characteristic samples with La, Ce, and Pr content exhibit an onset of Meissner screening in magnetic susceptibility measured in low fields, observable more clearly on the semi-log scale of the inset figure. 
Full volume fraction screening is not observed in these crystals, which exhibit partial screening estimated to reach as high as $\sim 10\%$ of full volume fraction, as shown for the case of 14\% Pr. While much smaller than that expected for a bulk superconducting material, this peculiar superconducting phase seems to be impervious to annealing, strong surface etching and surface oxidation, suggesting that the superconducting fraction of these samples is not obviously an extrinsic phase.

%%%%%%%%%%%%%%%%%%%%%%% Figure - RESISTIVITY

First, the persistent appearance of resistive transitions and diamagnetic screening signals indicating \tc\ values between 30~K and 47~K only occur in samples sufficiently doped such that the AFM phase is suppressed, as shown in Fig.~\ref{ressusc}. For instance, superconductivity with much lower \tc\ values often appears in very low-doped samples such as Ca$_{0.92}$La$_{0.08}$Fe$_2$As$_2$, which shows a resistive transition near $\sim 10$~K (Fig.~\ref{ressusc}a). This ``10~K phase'' persistently appears at low rare earth concentrations in the \CaR\ series, and often shows traces in resistivity data of pure \Ca\ such as shown in Fig.~\ref{resistivity}. The origin of this phase is not known, but is likely related to the strain-induced transition observed under non-hydrostatic pressure conditions in undoped \Ca. \cite{Torikachvili,Park,Prokes}
The high-\tc\ state induced by rare earth substitution appears to be a phase distinct from the 10~K phase, and occasionally exhibits distinct partial resistive transitions as displayed in Fig.~\ref{ressusc}(a) for 18\% La. The higher-$T_c$ phase in \CaR\ is likely stabilized by the extra carries introduced by aliovalent substitution, as isovalent substitution fails to induce superconductivity.\cite{Kasahara}

However, higher \tc\ transitions are never observed in resistivity for $x$ values below the concentrations necessary to suppress the AFM phase, ruling out any randomly occurring impurity or contaminant phase and suggesting that this pairing mechanism is strongly tied to the suppression of magnetism. It should also be noted that the presence of impurity or contaminant phases has not appeared in any scattering or chemical analysis experiments of these samples. In particular, single-crystal x-ray scattering data refinements of over twenty different crystals have not indicated the presence of any crystalline phases other than the 122 structure, and have consistently yielded residual fitting factors never greater than $\sim3\%$. More detailed (\ie, synchrotron) scattering experiments are required to further reduce the possible level of impurity phase that may be present, but the consistent absence of this phase in AFM samples statistically makes a strong case against this possibility.

%%%%%%%%% FIGURE:  SC Meissner screening 

Second, the magnitude of the \tc\ values observed is far above the transition temperatures that appear in undoped \Ca\ under non-hydrostatic pressure conditions (\ie, ``10~K'' phase) \cite{Torikachvili,Park,Prokes} or in undoped SrFe$_2$As$_2$ under strain (\ie, ``20~K'' phase),\cite{Saha_Sr} thus reducing the possibility of superconductivity due to strain conditions in parts of the sample. While we cannot rule out that strain is not playing any role whatsoever, the fact that the dramatic change in lattice constants through the collapsed structural transition does not influence the effectiveness of this potential strain mechanism is extremely challenging for such a scenario.

%%%%%%%%%% FIGURE:  ANNEALING/ETCHING
\begin{figure}[!t] \centering
  \resizebox{8.5cm}{!}{
  \includegraphics[width=8.5cm]{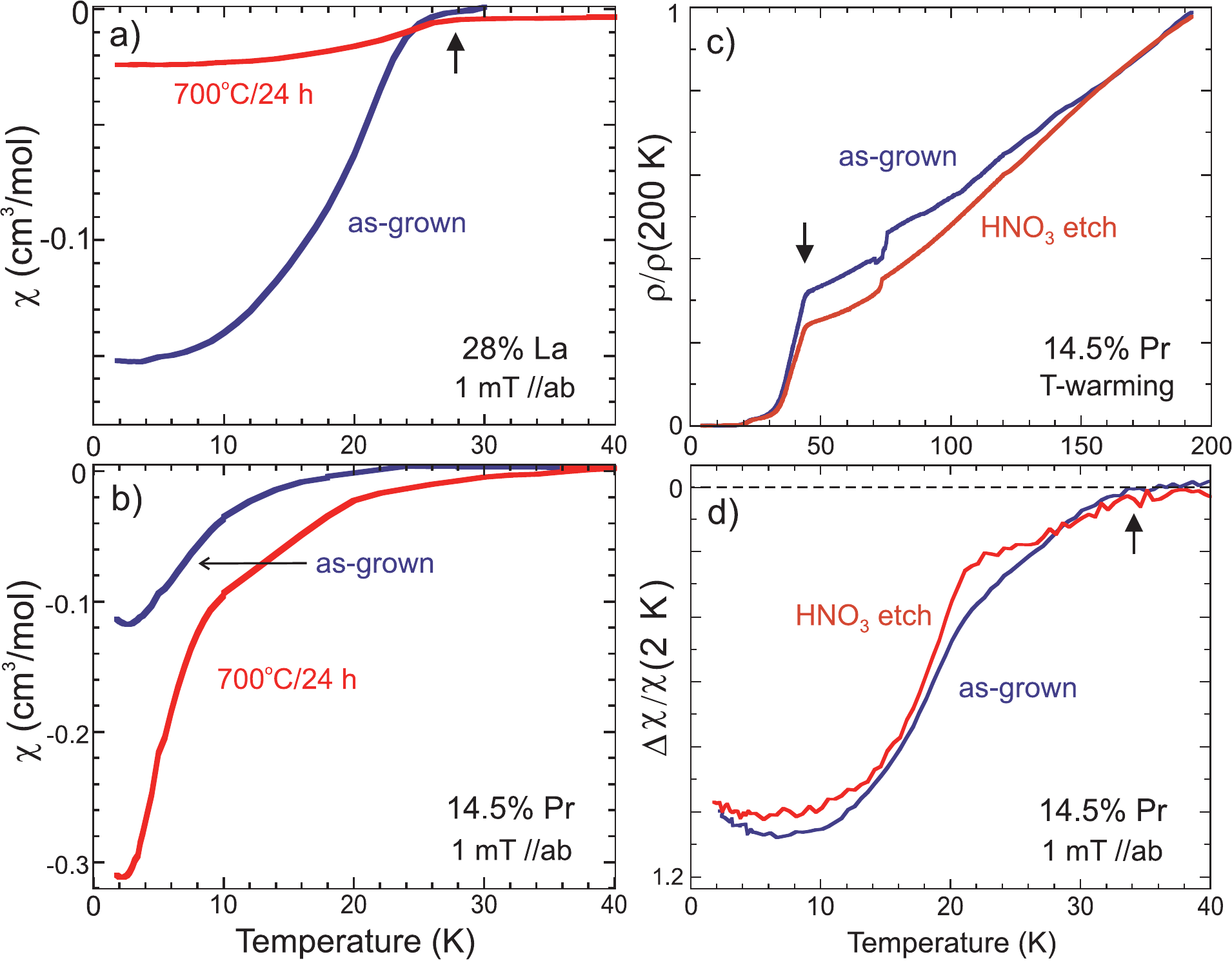}}
  \caption{\label{annealetch} Effects of annealing heat treatments on Meissner screening in non-collapsed (panel (a): La) and collapsed (panel (b): Pr) phases showing superconductivity. Effects of etching on (c) electrical resistivity (normalized to 200~K values) and (d) diamagnetic screening, represented by susceptibility offset from value above $T_c$ and normalized to 2~K value in a 15\% Pr-doped sample. Vertical arrows indicate position of onset of superconducting transition. All susceptibility data shown for zero-field-cooled conditions only for clarity.
  }
\end{figure}

Nevertheless, to further reduce strain mechanisms as a concern we have performed annealing studies of superconducting samples, both with and without structural collapse conditions present.
Starting with as-grown La- and Pr-substituted samples that exhibit Meissner screening, we first performed susceptibility measurements of each sample to characterize their as-grown properties and then subjected each sample to an annealing treatment. This consisted of sealing each sample in a separate quartz tube together with a Ta foil oxygen getter under partial Ar gas pressure, heating to 700$^{\circ}$C and holding at that temperature for 24 hrs before cooling to room temperature. Immediately after the annealing sequence, the susceptibility of each sample was measured following the same procedure as before.
Figs.~\ref{annealetch}(a)-(b) present the results of the before- and after-anneal measurements. Although there are finite changes in measured screening fractions, the main result is that both samples still exhibit Meissner screening after their annealing treatments. Furthermore, while the La-substituted sample shows a reduction in diamagnetic signal, the Pr-substituted sample in fact shows a small enhancement, reflective of the absence of any systematic trends to enhancing or reducing Meissner screening under this heat treatment schedule. This is in stark contrast to what happens in stoichoimetric \Sr, where annealing completely removes any signature of superconductivity.\cite{Saha_Sr}

%%%%%%%%%% FIGURE:  OXIDATION
\begin{figure}[!t] \centering
  \resizebox{8cm}{!}{
  \includegraphics[width=7.5cm]{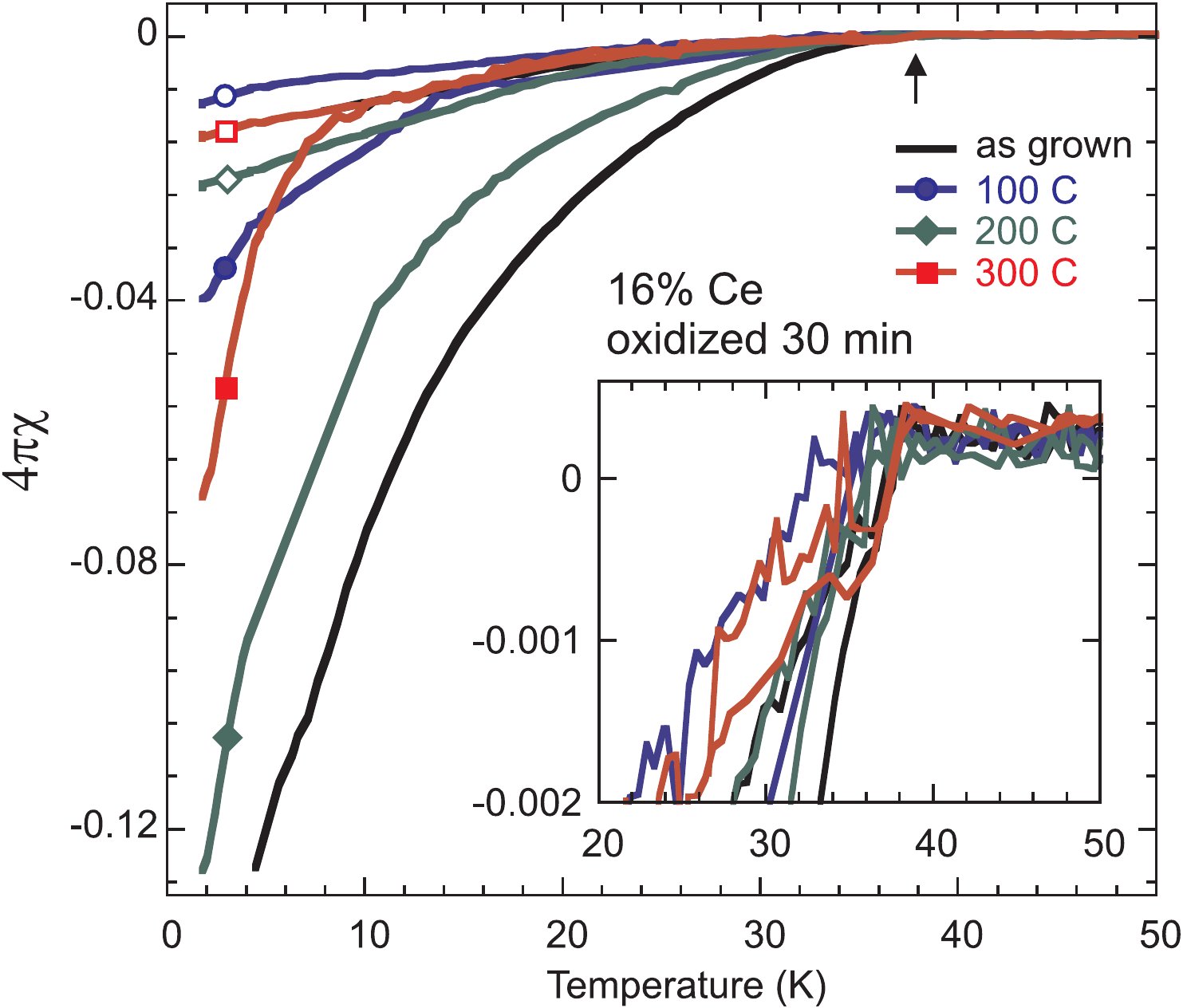}}
  \caption{\label{oxidize} Effect of surface oxidation on a Ca$_{0.84}$Ce$_{0.16}$Fe$_2$As$_2$ crystal, shown for the as-grown sample (black line) and for repeated exposures to air under heated conditions. Closed symbols indicate zero-field-cooled data, and open symbols indicate field-cooled data. Inset: zoom of main panel demonstrating the insensitivity of onset of Meissner screening at $T_c\simeq 37$~K to different heated exposures. }
\end{figure}

Finally, to rule out the possibility that surface impurity phases are responsible for partial Meissner screening, we have checked the effect of both etching and oxidation on the superconductivity in \CaR. With \tc\ values approaching those of the oxygen-based iron-pnictide superconductors, it is important to check for the possibility that oxygenated surface phases somehow achieve optimal oxygen doping for superconductivity and are providing the partial screening observed here. As shown in Figs.~\ref{annealetch}(c)-(d), we have measured both resistivity and magnetic susceptibility of a 14.5\% Pr sample both before and after etching the sample in concentrated HNO$_3$ for 30~s, which removes $\sim25\%$ of its mass. It is clear that superconductivity survives this harsh treatment, which results in no change in qualitative screening behavior, as well as very little change in resistivity signatures of both the collapse transition near 70~K and the superconducting transition that onsets at 40~K. 

To further verify that oxidation is not the cause of enhanced screening, the susceptibility of a 16\% Ce-doped sample with $T_c$=35~K was measured first as-grown and then after subsequent exposures to air under heated conditions on a temperature-controlled hot plate. As shown in Fig.~\ref{oxidize}, there is again no systematic trend observed after repeated oxidations, with onset of Meissner screening not changing significantly even after visible oxidation from 300$^{\circ}$C exposure. (The small volume fraction variations, which are non-monotonic with exposure temperature, are likely due to uncertainty in mass changes due to handling, as well as damage to the sample from oxidation). 

Together, these tests strongly reduce the likelihood of the observed high-$T_c$ superconducting phase in \CaR\ originating from extrinsic sources such as strain mechanisms, surface states or foreign phases such as oxides or other contaminants. However, assuming this superconducting phase is of the conventional type, the consistent observation of such small superconducting volume fractions points to a phase that does not occupy the bulk of the samples. This is extremely surprising, given that the majority of FeAs-based superconducting compounds exhibit bulk superconductivity upon suppression of the AFM phase.\cite{Paglione} We can speculate on its origin as having a localized nature tied to the low percentage of rare earth substitution, but further characterization is required to elucidate the origin of this phase and its potential to be stabilized in bulk form.

%%%%%%%%%%%%%%%%%%%%%%%%%%%%%%%%%%%%%%%%%%%%%%%%%%%%%%%%%%%%%%%%%%%%%%%%%%%%%%%%%%
\section{Phase Diagrams}

%%%%%%%%%% FIGURE:  phase diagrams
\begin{figure}[!t] \centering
  \resizebox{8cm}{!}{
  \includegraphics[width=8cm]{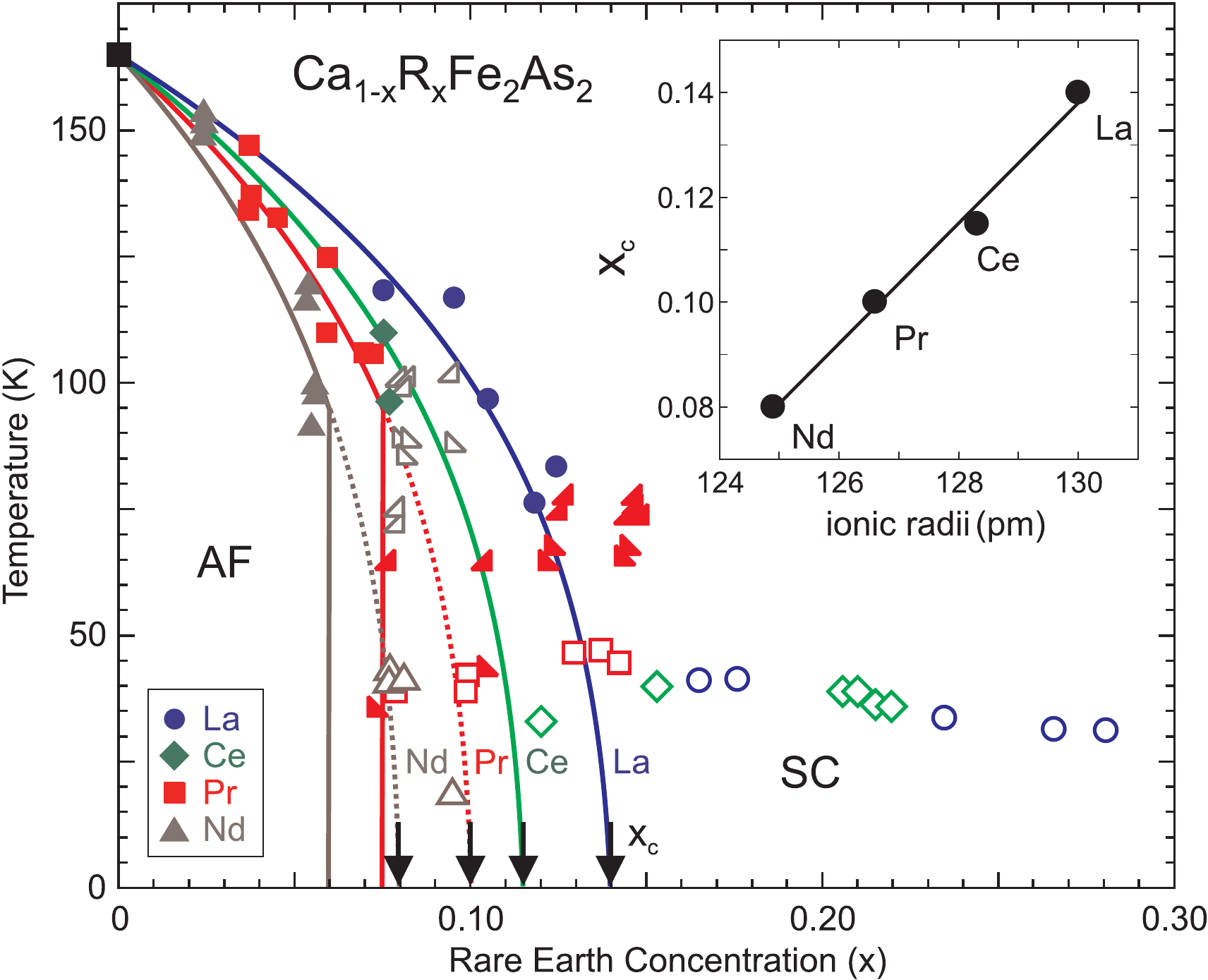}}
  \caption{\label{phasediagram} Rare earth substitution phase diagrams of \CaR, showing antiferromagnetic (AF) transitions (solid symbols), structural collapse transitions (half triangles) and superconducting transitions (open symbols) for Nd, Pr, Ce and La  substitutions (all concentrations determined by WDS; see text for method of determination). Half triangles indicate position of structural collapse transition on warming (right-pointing) and cooling (left-pointing) for Nd (open symbol) and Pr (closed symbol) substitutions. Solid lines are fits to AF transitions for each rare earth set, extrapolated to zero temperature to identify the critical concentration $x_c$ for each. Inset: scaling of $x_c$ with ionic radii of each rare earth species.}
\end{figure}

Because pressure \cite{Klintberg} and doping \cite{Kimber} are both effective in suppressing the AFM transition, the temperature-doping phase diagrams of the \CaR\ series appear qualitatively similar but in fact evolve with different concentration rates that depend on rare earth ionic size. Fig.~\ref{phasediagram} presents the composite phase diagrams of \CaR\ for each rare earth species, with antiferromagnetic transitions defined by the minimum in $\rho(T)$ and the midpoint of the drop in $\chi(T)$, superconducting transitions determined by the onset of a drop in $\rho(T)$, and CT transitions defined by abrupt features in $\chi(T)$ upon warming and cooling as discussed in Section~IV.

The suppression of the AFM phase with $x$ is similar for each species, but progresses at noticeably different rates. Extrapolating a phenomelogical fit of $T_N$ as a function of $x$ to $T=0$ shows this explicitly: the resultant critical concentration $x_{c}$ where $T_N$ vanishes is shown to vary with rare earth. In the inset of Fig.~\ref{phasediagram}, we show that $x_{c}$ actually scales linearly with the ionic radii values of the rare earth species for 8-coordinate geometry.\cite{Shannon} Given the known sensitivity of the lattice parameters to choice of rare earth substituent as shown by the structural characterization in Section~III, this trend verifies that, in addition to electron doping, chemical pressure also plays a role in shaping the phase diagram of the \CaR\ system.

To disentangle the doping and pressure effects, we utilize the observations noted in Section~III about the progression of lattice constants -- in particular the strong and weak dependences of $c$- and $a$-axis lattice constants, respectively, on rare earth species (c.f. see Fig.~\ref{xray1}) -- to characterize chemical pressure by the measured change in $c$-axis unit cell dimension. For instance, substitution of La into \Ca\ does not change the $c$-axis unit cell length for concentrations up to almost 30\% La, while Nd substitution changes the $c$-axis very rapidly with $x$. However, for all rare earth species the $a$-axis length increases on average at the same rate with substitution concentration regardless of ionic size. This is possibly due to an expansion of the Fe sublattice caused by charge doping with an effective adjustment of the Fe oxidation state, but such a conclusion requires verification from a core level spectroscopy experiment.

%%%%%%%%%% FIGURE:  phase diagrams
\begin{figure}[!t]  \centering
  \resizebox{8cm}{!}{
  \includegraphics[width=8cm]{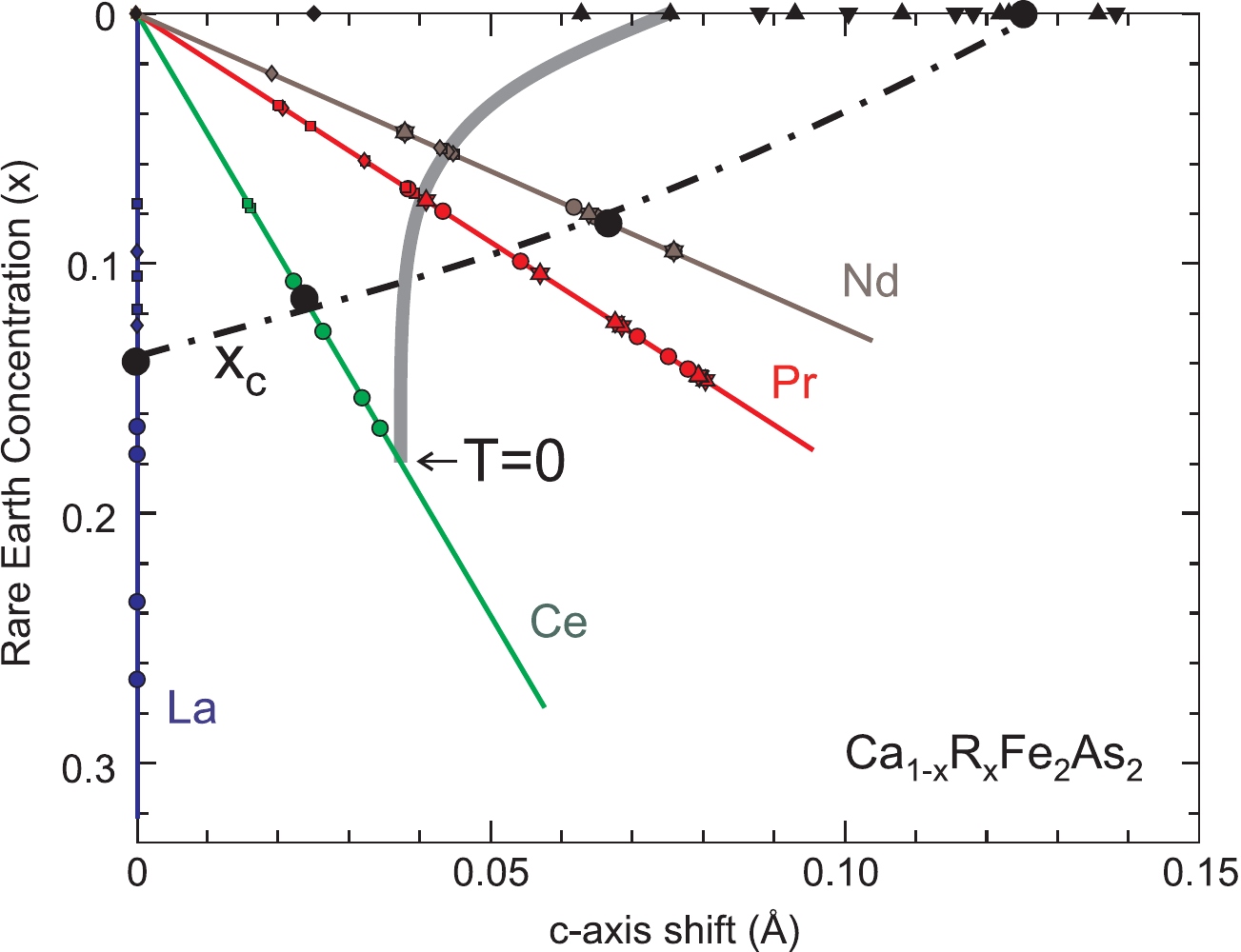}}
  \caption{\label{projection} The effective electronic doping-pressure phase diagram for \CaR\ and undoped \Ca\ under pressure (data from Ref.~\onlinecite{Goldman}) used to construct the phase diagram of Fig.~\ref{phasediag}. The individual rare earth series are projections of finite temperature data onto the $T=0$ plane, showing the separation of the effects of electron doping ($x$-axis) and chemical pressure ($c$-axis shift).  
  The dash-dotted line is a guide to $x_c$ critical points denoting the extrapolated suppression of the antiferromagnetic phase to $T=0$. The solid grey line indicates the projection of the doping-pressure position where the As-As interlayer distance equals 3~\AA.}
\end{figure}

%%%%%%%%%%%%%%%%%%%%%%% Figure - PHASE DIAGRAM
\begin{figure*}[!t] \centering
  \resizebox{15cm}{!}{
  \includegraphics[width=9cm]{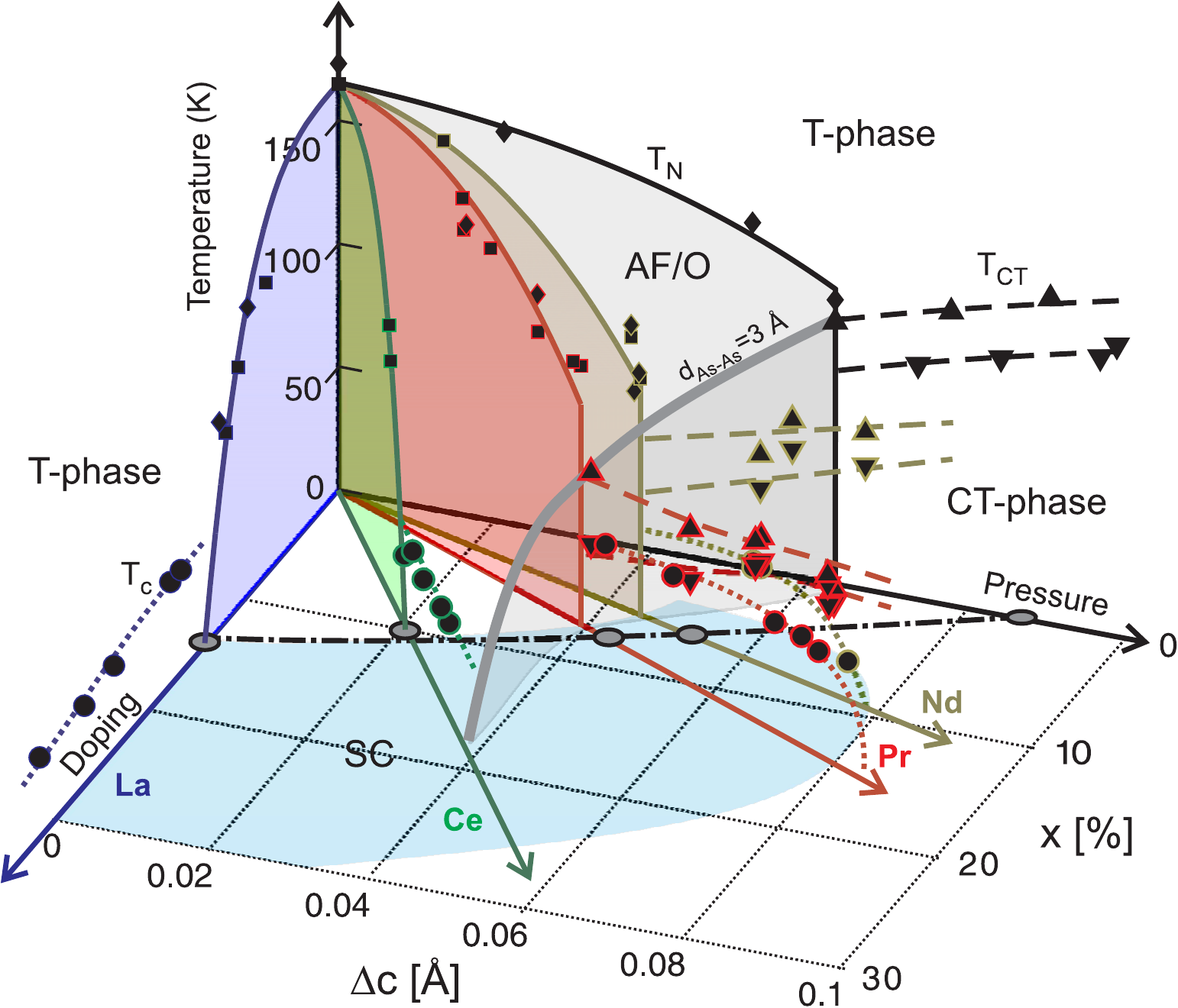}}
  \caption{\label{phasediag} Phase diagram of the \CaR\ series showing the evolution of the antiferromagnetic transition $T_N$, the appearance of superconductivity at \tc,  and the isostructural collapse as a function of electron doping ($x$) and effective chemical pressure ($\Delta c$, the measured change in $c$-axis induced by doping relative to the value of undoped \Ca\ at ambient pressure). Data for $x=0$ are taken from \Ca\ measurements under pressure. \cite{Goldman} Data points for the AFM transition are obtained from electrical resistivity (diamonds) and magnetic susceptibility (squares). Superconducting transitions are taken from resistivity data (circles), and collapse transitions (triangles) are from susceptibility data, indicating warming (up-triangle) and cooling (down-triangle) conditions. The solid grey line indicates the position where the interlayer As-As separation equals 3~\AA, coinciding with the onset of the structural collapse for each rare earth series and for \Ca\ under pressure. The blue shaded $T=0$ plane indicates the range where superconductivity is observed.
  }
\end{figure*}

Therefore, we take the change in $c$-axis length as a measure of the true chemical pressure. The value of the $c$-axis lattice parameter at $x_{c}$ for each $R$ is then used to project the individual phase diagrams onto the $x$--$c$-axis plane, as shown in Fig.~\ref{projection}. Note the smooth progression of the $x_c$ points across this plane (\ie, dash-dotted line), which also includes the same extrapolated critical point for undoped \Ca\ under pressure.\cite{Goldman} This indicates that a parameterized relation exists between the suppression of AFM order via chemical pressure and electron doping.

Also, the positions where the As-As interlayer distance equals 3~\AA\ are shown as a solid grey line, indicating the positions where the collapse transition onsets. This construction forms the basis for the universal phase diagram for \CaR\ shown in Fig.~\ref{phasediag}, which extends the pressure-temperature phase diagram of \Ca\ (Refs.~\onlinecite{Torikachvili,Kreyssig,Yu}) along a third charge-doping axis. In this manner, it is seen that the individual rare earth species phase diagrams nicely straddle the doping-pressure plane in a manner that allows access to distinct parts of the phase diagram by choice of rare earth substituent.

%%%%%%%%%%%%%%%%%%%%%%%%%%%%%%%%%%%%%%%%%%%%%%%%%%%%%%%%%%%%%%%%%%%%%%%%%%%%%%%%%%
\section{Discussion and Summary}

Combining our transport, magnetic susceptibility and neutron scattering data enables us to trace the progression of the AFM, CT and superconducting transitions as a function of the segregated parameters of electron doping and chemical pressure.
In this way, it can be seen that the AFM phase is effectively suppressed by both doping and chemical pressure, similar to other established systems such
as Ba$_{1-x}$K$_x$Fe$_2$As$_2$ and pressurized BaFe$_2$As$_2$.\cite{Kimber} Furthermore, the AFM transition line exhibits continuity through the doping-pressure plane, demonstrating the symmetry between both methods of tuning. This is in line with ideas about band structure tuning, whereby nesting features of the Fermi surface that may stabilize the AFM phase in the parent compound are disrupted by either tuning parameter. However, the suppression of AFM order with electron doping at the alkaline earth site is in stark contrast to recent first-principles calculations that predict an enhancement of magnetism,\cite{Johannes} demonstrating the failure of a rigid band picture even at low charge doping. This is reflective of a continued challenge to understand the true nature of the magnetic order and its suppression by doping and pressure.

As in undoped \Ca\ under pressure,\cite{Yu,Goldman} where the CT phase abruptly severs the continuous suppression of the AFM transition under applied pressure, the suppression of the AFM phase in \PrCa\ and \NdCa\ is also shown to be interrupted by the CT transition but at slightly lower temperatures and effective pressures. This is understood as due to the occurrence of the CT transition exactly at the 3~\AA\ interlayer As-As separation, which follows both a pressure- and doping-dependent path through the phase diagram as marked by the solid grey line in Fig.~\ref{phasediag}. What is more unusual is the insensitivity of the observed high-\tc\ superconducting phase to this boundary, raising important questions regarding which elements of chemical, electronic and magnetic structure are important to Cooper pairing should this superconducting phase prove to be intrinsic to both the uncollapsed and collapsed structures that straddle this division.

Interestingly, high-temperature superconductivity in the \CaR\ series appears to exist only exclusively from the AFM phase. This is strikingly similar to the segregation of SC and AFM phases found in 1111 materials doped with fluorine, such as in LaFeAsO$_{1-x}$F$_x$ (Ref.~\onlinecite{Luetkens}) and CeFeAsO$_{1-x}$F$_x$ (Ref.~\onlinecite{Zhao}), and should be contrasted with the well-known coexistence shown to occur in BaFe$_{2-x}$Co$_x$As$_2$. \cite{Ni,Chu} Further confirmation of the intrinsic nature of superconductivity in \CaR\ will shed light on this interesting dichotomy, possibly providing an explanation for this distinction between phase diagrams in oxypnictide-based and intermetallic-based superconductors.

In conclusion, we have shown that rare earth substitution into the iron-based superconductor parent compound \Ca\ provides for a rich playground of phases that will prove useful for studying various aspects of the physics of iron-based superconductivity. Depending on the choice of rare earth substituent, varying degrees of chemical pressure and electron doping can be utilized to tune both the electronic and structural phases of this system, resulting in a remarkable control over a large phase space of temperature, doping and pressure. 

We have shown that chemical pressure can drive \Ca\ through a structural collapse of the tetragonal unit cell that retains the crystal symmetry, but dramatically changes the bonding structure, dimensionality and electronic properties. The collapse is driven solely by the interlayer As-As $p$-orbital separation, which prefers to form a covalent bond when the separation is driven to less than 3~\AA\ by chemical substitution or applied pressure, or a combination of both. This results in an unprecedented thermal expansion of the unit cell due to this instability, and a controllable tunability of the crystal and electronic structure as a function of temperature. 

Interestingly, an unprecedentedly high superconducting transition temperature was observed in all rare earth substitutions upon complete suppression of the antiferromagnetically ordered phase, with several extrinsic origins of this partial-volume-fraction phase systematically ruled out. The presence of this superconductivity regardless of the structural collapse instability raises important questions regarding the sensitivity of Cooper pairing in the iron-based materials to electronic structure, bonding and dimensionality, and access to this dramatic structural collapse at ambient pressure conditions will provide ample opportunity to study these effects in further detail.

%%%%%%%%%%%%%%%%%%%%%%%%%%%% ACKNOWLEDGMENTS
\section{Acknowledgements}

The authors gratefully acknowledge B.~Eichhorn, M.A.~Green, R.L.~Greene, I.I.~Mazin, J.~Schmalian and I.~Takeuchi. This work was supported by AFOSR-MURI Grant FA9550-09-1-0603 and NSF-CAREER Grant DMR-0952716.

%%%%%%%%%%%%%%%%%% BIBLIOGRAPHY %%%%%%%%%%%%%%%%%%%%


\begin{thebibliography}{11}


\bibitem{Kamihara} Y. Kamihara, T. Watanabe, M. Hirano, and H. Hosono,
%Iron-based layered superconductor La[O$_{1-x}$F$_x$]FeAs ($x$=0.05-0.12) with T$_c$=26 K.
J. Am. Chem. Soc. {\bf 130}, 3296 (2008).

\bibitem{Paglione} J. Paglione and R. L. Greene, 
%High-temperature superconductivity in iron-based materials. 
Nature Phys. {\bf 6}, 645 (2010).

\bibitem{Just} G. Just and P. Paufler, 
%On the coordination of ThCr$_2$Si$_2$ (BaAl$_4$)-type compounds within the field of free parameters. 
J. Alloys Comp. {\bf 232}, 1-25 (1996).

\bibitem{Hoffman} R. Hoffmann and C. Zheng, 
%Making and breaking bonds in the solid state: the thorium chromium silicide (ThCr$_2$Si$_2$) structure. 
J. Phys. Chem. {\bf 89}, 4175 (1985).

\bibitem{SahaSCES} S. R. Saha, K. Kirshenbaum, N. P. Butch, J. Paglione, and P. Y. Zavalij,   
%Uniform chemical pressure effect in solid solutions Ba$_{1-x}$Sr$_x$Fe$_2$As$_2$ and Sr$_{1-x}$Ca$_x$Fe$_2$As$_2$. 
J. Phys.: Conf. Ser. {\bf 273}, 012104 (2011).

\bibitem{Kreyssig} A. Kreyssig, M. A. Green, Y. Lee, G. D. Samolyuk, P. Zajdel, J. W. Lynn, S. L. Budko, M. S. Torikachvili, N. Ni, S. Nandi, J. B. Leao, S. J. Poulton, D. N. Argyriou, B. N. Harmon, R. J. McQueeney, P. C. Canfield, and A. I. Goldman, 
%Pressure-induced volume-collapsed tetragonal phase of CaFe$_2$As$_2$ as seen via neutron scattering. 
Phys. Rev. B {\bf 78}, 184517 (2008).

\bibitem{Goldman} A. I. Goldman, A. Kreyssig, K. Proke, D. K. Pratt, D. N. Argyriou, J. W. Lynn, S. Nandi, S. A. J. Kimber, Y. Chen, Y. B. Lee, G. Samolyuk, J. B. Leao, S. J. Poulton, S. L. Budko, N. Ni, P. C. Canfield, B. N. Harmon, and R. J. McQueeney, 
%Lattice collapse and quenching of magnetism in CaFe$_2$As$_2$ under pressure: a single-crystal neutron and x-ray diffraction investigation. 
Phys. Rev. B {\bf 79}, 024513 (2009).

\bibitem{Torikachvili}M. S. Torikachvili, S. L. Budko, N. Ni, and P.C. Canfield, 
%Pressure-induced volume-collapsed tetragonal phase of CaFe$_2$As$_2$. 
Phys. Rev. Lett. {\bf 101}, 057006 (2008).

%Pressure-induced superconductivity in CaFe2As2
\bibitem{Park} T. Park, E. Park, H. Lee, T. Klimczuk, E. D. Bauer, F. Ronning, and J. D. Thompson,  J. Phys.: Cond. Matter {\bf 20}, 322204 (2008).

%SC stabilized in tetragonal phase of \Ca\ under uniaxial pressure
\bibitem{Prokes} K. Prokes, A. Kreyssig, B. Ouladdiaf, D. K. Pratt, N. Ni, S. L. Budko, P. C. Canfield, R. J. McQueeney, D. N. Argyriou, and A. I. Goldman, 
Phys. Rev. B {\bf 81}, 180506R (2010).

\bibitem{Yu} W. Yu, A. A. Aczel, T. J. Williams, S. L. Budko, N. Ni, P. C. Canfield, and G. M. Luke,  
%Absence of superconductivity in single-phase CaFe$_2$As$_2$ under hydrostatic pressure. 
Phys. Rev. B {\bf 79}, 020511R (2009).

%Abrupt recovery of Fermi-liquid transport following the collapse of the $c$ axis in CaFe$_2$(As$_{1-x}$P$_x$)$_2$ single crystals.
\bibitem{Kasahara} S. Kasahara, T. Shibauchi, K. Hashimoto, Y. Nakai, H. Ikeda, T. Terashima, and Y. Matsuda, Phys. Rev. B {\bf 83}, 060505R (2011).

\bibitem{Yildirim} T. Yildirim, 
%Strong coupling of the Fe-spin state and the As-As hybridization in iron-pnictide superconductors from first-principle calculations. 
Phys. Rev. Lett. {\bf 102}, 037003 (2009).

\bibitem{Paglione_APS} J. Paglione, BAPS.2011.MAR.W26.4; S. R. Saha \etal, arXiv:1105.4798 (2011).

\bibitem{Shannon} R. D. Shannon, 
%Revised effective ionic radii and systematic studies of interatomic distances in halides and chalcogenides. 
Acta Cryst. {\bf A32}, 751-767 (1976).

\bibitem{Saha_Sr} S. R. Saha, N. P. Butch, K. Kirshenbaum, J. Paglione, and P. Y. Zavalij, 
%Superconducting and ferromagnetic phases induced by lattice distortions in stoichiometric SrFe$_2$As$_2$ single crystals. 
Phys. Rev. Lett. {\bf 103}, 037005 (2009).

\bibitem{Muraba} Y. Muraba, S. Matsuishi, S-W. Kim, T. Atou, O. Fukunaga, and H. Hosono, 
%High-pressure synthesis of the indirectly electron-doped iron pnictide superconductor Sr$_{1-x}$La$_x$Fe$_2$As$_2$ with maximum $T_c$=22 K. 
Phys. Rev. B {\bf 82}, 180512R (2010).

\bibitem{Sheldrick}G. M. Sheldrick, Acta Cryst. {\bf A64}, 112-122 (2008).

%% REF IN FIG.5 %%
\bibitem{Kumar_Ni} N. Kumar, R. Nagalakshmi, R. Kulkarni, P. L. Paulose, A. K. Nigam, S. K. Dhar, and A. Thamizhavel, 
%Anisotropic magnetic and superconducting properties of CaFe$_{2-x}$Co$_x$As$_2$ ($x$=0, 0.06) single crystals. 
Phys. Rev. B {\bf 79}, 012504 (2009).

\bibitem{Ran-Canfield} S. Ran, S. L. Budko, D. K. Pratt, A. Kreyssig, M. G. Kim, M. J. Kramer, D. H. Ryan, W. N. Rowan-Weetaluktuk, Y. Furukawa, B. Roy, A. I. Goldman, and P. C. Canfield, Phys. Rev. B {\bf 83}, 144517 (2011).

\bibitem{CRC} CRC Handbook of Chemistry and Physics, edited by D.R. Lide (CRC Press, Boca Raton, FL, 2007), 88th ed.

\bibitem{Das} D. Das, T. Jacobs, and L. J. Barbour,  
%Exceptionally large positive and negative anisotropic thermal expansion of an organic crystalline material. 
Nature Mat. {\bf 9}, 36 (2010). 

\bibitem{Huhnt} C. Huhnt, G. Michels, M. Roepke, W. Schlabitz, A. Wurth, D. Johrendt, and A. Mewis, 
%First-order phase transitions in the ThCr$_2$Si$_2$-type phosphides ARh$_2$P$_2$ (A = Sr, Eu). 
Physica B {\bf 240}, 26-37 (1997).

\bibitem{Ronning} F. Ronning, T Klimczuk, E. D. Bauer, H. Volz, and J. D. Thompson, 
J. Phys.: Condens. Matt. {\bf 20}, 322201 (2008).

\bibitem{Ni78} N. Ni, S. Nandi, A. Kreyssig, A. I. Goldman, E. D. Mun, S. L. Budko, and P. C. Canfield, Phys. Rev. B {\bf 78}, 014523 (2008).

\bibitem{Goldman78} A. I. Goldman, D. N. Argyriou, B. Ouladdiaf, T. Chatterji, A. Kreyssig, S. Nandi, N. Ni, S. L. Budko, P. C. Canfield, and R. J. McQueeney, Phys. Rev. B {\bf 78}, 100506 (2008).

\bibitem{Rotter} M. Rotter, M. Tegel, and D. Johrend,  
%Superconductivity at 38 K in the Iron Arsenide (Ba$_{1-x}$K$_x$)Fe$_2$As$_2$. 
Phys. Rev. Lett. {\bf 101}, 107006 (2008).

\bibitem{Cheng} P. Cheng, B. Shen, G. Mu, X. Zhu, F. Han, B. Zeng and H. -H. Wen,  
%High-Tc superconductivity induced by doping rare-earth elements into CaFeAsF. 
Europhys. Lett. {\bf 85}, 67003 (2009).

\bibitem{Klintberg} L.E. Klintberg, S. K. Goh, S. Kasahara, Y. Nakai, K. Ishida, M. Sutherland, T. Shibauchi, Y. Matsuda, and T. Terashima, 
%Chemical Pressure and Physical Pressure in BaFe$_2$(As$_{1-x}$P$_x$)$_2$. 
J. Phys. Soc. Japan {\bf 79}, 123706 (2010).

\bibitem{Kimber} S.A. J. Kimber, A. Kreyssig, Y.-Z. Zhang, H. O. Jeschke, R. Valenti, F. Yokaichiya, E. Colombier, J. Yan, T. C. Hansen, T. Chatterji, R. J. McQueeney, P. C. Canfield, A. I. Goldman, and D. N. Argyriou, , 
%Similarities between structural distortions under pressure and chemical doping in superconducting BaFe$_2$As$_2$. 
Nature Mat. {\bf 8}, 471 (2009).


\bibitem{Johannes} M.D. Johannes, I.I. Mazin, and D. S. Parker, 
%Microscopic origin of magnetism and magnetic interactions in ferropnictides. 
Phys. Rev. B {\bf 82}, 024527 (2010).

\bibitem{Luetkens} H. Luetkens, H.-H. Klauss, M. Kraken, F. J. Litterst, T. Dellmann, R. Klingeler, C. Hess, R. Khasanov, A. Amato, C. Baines, M. Kosmala, O. J. Schumann, M. Braden, J. Hamann-Borrero, N. Leps, A. Kondrat, G. Behr, J. Werner, and B. Büchner,  
%The electronic phase diagram of the LaO$_{1-x}$F$_x$FeAs superconductor. 
Nature Mat. {\bf 8}, 305 (2009).

\bibitem{Zhao} J. Zhao, Q. Huang, C. d. l. Cruz, S. Li, J. W. Lynn, Y. Chen, M. A. Green, G. F. Chen, G. Li, Z. Li, J. L. Luo, N. L. Wang, and P. Dai, , 
%Structural and magnetic phase diagram of CeFeAsO$_{1-x}$F$_x$ and its relation to high-temperature superconductivity. 
Nature Mat. {\bf 7}, 953 (2008).

\bibitem{Ni} N. Ni, M. E. Tillman, J.-Q. Yan, A. Kracher, S. T. Hannahs, S. L. Budko, and P. C. Canfield,  
%Effects of Co substitution on thermodynamic and transport properties and anisotropic $H_{c2}$ in Ba(Fe$_{1-x}$Co$_x$)$_2$As$_2$ single crystals. 
Phys. Rev. B {\bf 78}, 214515 (2008).

\bibitem{Chu} J.-H. Chu, J. G. Analytis, C. Kucharczyk, and I. R. Fisher,  
%Determination of the phase diagram of the electron-doped superconductor Ba(Fe$_{1-x}$Co$_x$)$_2$As$_2$. 
Phys. Rev. B {\bf 79}, 014506 (2009).




\end{thebibliography}
\end{document}